\documentclass[11pt,english]{article}

\usepackage[utf8]{inputenc} 

\usepackage{dsfont}

\usepackage[left=1in,right=1in,top=1in,bottom=1in]{geometry}
\usepackage{authblk}
\usepackage{enumitem}
\usepackage{ulem} 
\usepackage{empheq}

\usepackage{graphicx}
\usepackage{subcaption}
\usepackage{float}
\usepackage{wrapfig} 

\usepackage{xcolor}
\usepackage{color}
\definecolor{darkgreen}{rgb}{0,0.5,0}
\definecolor{darkblue}{rgb}{0,0,0.6}
\definecolor{purple}{rgb}{0.4,.2,0.7}

\usepackage{amsmath}
\usepackage{amssymb}

\usepackage{tensor} 
\usepackage{slashed}
\usepackage{mathtools}
\usepackage{leftidx}
\usepackage{esint}
\usepackage{amsfonts,amsthm,bm}

\usepackage{physics}
\usepackage{feynmf}
\usepackage{braket}
\usepackage{caption}

\usepackage{prettyref}
\usepackage[colorlinks=true,citecolor=darkgreen,linkcolor=purple,urlcolor=purple]{hyperref}

\newcommand{\nn}{\nonumber}

\numberwithin{equation}{section}
\numberwithin{figure}{section}
\numberwithin{table}{section}

\begin{document}

\title{\LARGE\textsc{Black Hole Scattering and Partition Functions}}  


\author[a]{\vskip1cm \normalsize Y.T. Albert Law}
\affil[a]{ \it \normalsize	 Center for the Fundamental Laws of Nature, Harvard University, Cambridge, MA 02138, USA}
\author[b]{Klaas Parmentier}
\affil[b]{ \it  Center for Theoretical Physics, Columbia University, New York, NY 10027, USA}
\date{}
\maketitle

\begin{center}
	\vskip-10mm
	{\footnotesize  \href{mailto:ylaw1@g.harvard.edu}{ylaw1@g.harvard.edu}\, ,\;  \href{mailto:k.parmentier@columbia.edu}{k.parmentier@columbia.edu}}
\end{center}

\vskip10mm

\thispagestyle{empty}


    

\begin{abstract}
 When computing the ideal gas thermal canonical partition function for a scalar outside a black hole horizon, one encounters the divergent single-particle density of states (DOS) due to the continuous nature of the normal mode spectrum. Recasting the Lorentzian field equation into an effective 1D scattering problem, we argue that the scattering phases encode non-trivial information about the DOS and can be extracted by  ``renormalizing" the DOS with respect to a reference. This defines a renormalized free energy up to an arbitrary additive constant. Interestingly, we discover that the 1-loop Euclidean path integral, as computed by the Denef-Hartnoll-Sachdev formula, fixes the reference free energy to be that on a Rindler-like region, and the renormalized DOS captures the quasinormal modes for the scalar. We support these claims with the examples of scalars on static BTZ, Nariai black holes and the de Sitter static patch. For black holes in asymptotically flat space, the renormalized DOS is captured by the phase of the transmission coefficient whose magnitude squared is the greybody factor. We comment on possible connections with recent works from an algebraic point of view.


\end{abstract}


\newpage

\tableofcontents



\section{Introduction}

Euclidean gravity methods \cite{Gibbons:1976ue} have been incredibly successful as an IR window into black hole microstates, even beyond the leading order in $G_N$. For example, as demonstrated in \cite{Banerjee:2010qc, Banerjee:2011jp, Sen:2012dw, Sen:2012kpz, Sen:2014aja}, 1-loop Euclidean path integrals compute logarithmic corrections to black hole entropy that are in perfect agreement with the microscopic results in string theory or holographic CFT. At 1-loop, the path integral receives corrections from quadratic fluctuations of matter fields and the graviton around the black hole, and reduces to functional determinants of differential operators. For a real scalar, 
\begin{align}\label{intro:scalar PI}
	Z_\text{PI}(m^2)=\int\mathcal{D}\phi \, e^{-\frac{1}{2}\int \left(\nabla \phi \right)^2 + m^2 \phi^2}=\frac{1}{\det \left(-\nabla^2+m^2 \right)^{1/2}} \; .
\end{align}

What is the {\it Lorentzian} interpretation of 1-loop path integrals such as \eqref{intro:scalar PI}? In other words, what is the computation in the canonical formalism that reproduces \eqref{intro:scalar PI}? In the case of a static spherically symmetric black hole, the path integral is periodic in Euclidean time, so a first thought would be that \eqref{intro:scalar PI} is equal to 
\begin{align}\label{intro:canfn}
    Z_\text{bulk}=\Tr e^{-\beta_H \hat H} \; ,
\end{align}
the canonical partition function for the free scalar outside the horizon at the inverse black hole temperature $\beta_H$. Here $\hat{H}$ is the Hamiltonian generating time translation, with respect to which one defines the creation and annihilation operators associated with the negative- and positive-energy normal modes respectively; Tr traces over the resulting Fock space. 

Were the spectrum of $\hat{H}$ discrete, one could compute \eqref{intro:canfn} by substituting the mode expansion for the scalar field and summing over the occupation numbers, leading to 
\begin{align}\label{intro:Zbulkdis}
    Z_\text{bulk}= \prod_{E>0} \frac{e^{-\beta_H E/2}}{1-e^{-\beta_H E}} \; .
\end{align}
Here the product is over the discrete {\it single-particle} energy spectrum labeled by $E$; the factor $e^{-\beta_H E/2}$ is due to the zero-point energy for each positive-energy mode. Equivalently, we can write
\begin{align}\label{intro:log Z bulk}
	\log Z_\text{bulk} = -\int_0^\infty d\omega \, \rho(\omega) \log \left( e^{\beta_H\omega/2}- e^{-\beta_H \omega /2}\right)  \; ,
\end{align}
in terms of the single-particle density of states (DOS)
\begin{align}\label{intro:dosdis}
    \rho(\omega) = \sum_{E>0} \delta(\omega -E) = \tr \delta(\omega -\hat H) 
\end{align}
with tr tracing over the single-particle Hilbert space.
For the case at hand, however, the spectrum of $\hat H$ is continuous; physically this is related to the fact that the horizon is an infinite redshift surface, enabling
the existence of normal modes with arbitrary angular momentum and energy. Subsequently, expressions \eqref{intro:Zbulkdis}-\eqref{intro:dosdis} do not make sense. The most common approach was suggested by 't Hooft \cite{tHooft:1984kcu}, where one discretizes the spectrum by putting a brick wall near the horizon and imposing a boundary condition (variants are reviewed in \cite{Frolov:1998vs,Solodukhin:2011gn}).
 In this way, one could apply formulas \eqref{intro:Zbulkdis}-\eqref{intro:dosdis}; at the end the answer depends on the brick wall parameter. However, this cannot be equal to the Euclidean path integral \eqref{intro:scalar PI}, which is manifestly covariantly defined.


Recently, motivated by constructing 1-loop tests for microscopic models of de Sitter quantum gravity, the authors of \cite{Anninos:2020hfj} studied 1-loop sphere path integrals. For scalars and spinors on $S^{d+1}$,
\begin{align}\label{introeq:DHSintegral}
\log Z_\text{PI} =\int_0^\infty \frac{dt}{2t}\left( \frac{1+e^{-2\pi t/\beta_\text{dS}}}{1-e^{-2\pi t/\beta_\text{dS}}}\,\chi^\text{scalar}(t)-\frac{2\, e^{-\pi t/\beta_\text{dS}}}{1-e^{-2\pi t/\beta_\text{dS}}}\chi^\text{spinor}(t)\right) \; .
\end{align}
Here $\beta_\text{dS}=2\pi \ell_\text{dS}$ is the inverse de Sitter temperature. By $\chi^\text{scalar}(t)$ and $\chi^\text{spinor}(t)$ we denote the Harish-Chandra characters of the de Sitter group $SO(1,d+1)$ for the scalars and spinors, which encode the quasinormal mode (QNM) spectrum on a static patch in $dS_{d+1}$:
\begin{align}\label{intro:qnmchar}
    \chi(t)= \sum_z N_z \, e^{-iz t}  \; .
\end{align}
Here $z$ and $N_z$ are the frequencies and degeneracies of the QNMs. An important observation from \cite{Anninos:2020hfj} is that if one {\it replaces} in \eqref{intro:log Z bulk}
\begin{align}
    \rho(\omega) \to \int_{0}^\infty \frac{dt}{2\pi} \left(e^{i \omega t} +e^{-i \omega t} \right)\, \chi(t) 
\end{align}
and takes $\beta_H=\beta_\text{dS}$, one recovers the scalar part of the sphere path integral \eqref{introeq:DHSintegral}. The same is true for spinors. The goal of this work is to elaborate on the physics of this replacement and extend to general static spherically symmetric black hole backgrounds.


Our starting point is to recast the problem into that of 1D scattering, at which \cite{Anninos:2020hfj} has already hinted. At each angular momentum $l\geq 0$, normal modes for the free scalar are equivalent to the scattering modes for the scattering problem. From this viewpoint, the continuum of the normal mode spectrum is identical in nature to that of scattering modes in any infinite-volume system with a localized potential. Because of this continuum, within any small interval $\Delta\omega$ of energy there are infinitely many scattering modes, and thus the density of states $\rho_l(\omega)$ is strictly infinite. 

A common strategy to extract useful spectral information for infinite-volume systems is to consider {\it changes} in the DOS upon changing the potential (see for instance \cite{ChaosBook}). For example, one can compare the original system to a reference system of a free particle (whose scattering problem has an exactly zero potential); the {\it difference} in the DOS and the resulting thermodynamic quantities then measure the effects due to the presence of a potential. Of course, in principle one is not restricted to choosing free particle as the reference system. This is the strategy we will pursue in this paper: while the DOS $\rho_l(\omega)$ is infinite, we can measure its difference from {\it some} reference DOS $\bar\rho_l(\omega)$; such a change is completely finite and is captured by the scattering phases associated with the corresponding scattering problems. Upon summing over $l\geq 0$, we thus have a manifestly covariant quantity $\log Z_\text{bulk}-\log \bar Z_\text{bulk}$, up to a choice of $\bar Z_\text{bulk}$.



Strikingly, we find that 1-loop Euclidean partition function \eqref{intro:scalar PI} uniquely fixes a reference $\bar Z_\text{bulk}$. To that end, we note that a formula for 1-loop determinants developed by Denef, Hartnoll and Sachdev (DHS) \cite{Denef:2009kn}, in terms of QNM frequencies of scalar and spinor fields on the black hole, can in fact be brought into the form \eqref{introeq:DHSintegral}, with $\beta_\text{dS}$ replaced by the inverse black hole temperature $\beta_H$ and the $SO(1,d+1)$ character $\chi(t)$ by a ``QNM character" defined as a sum analogous to \eqref{intro:qnmchar}. Comparing this with the above-mentioned Lorentzian computation for the examples of scalars on static BTZ, Nariai spacetime and the de Sitter static patch \cite{Anninos:2020hfj}, our central observation is that
\begin{align}\label{introeq:PI}
    Z_\text{PI}  = \widetilde{Z}_\text{bulk} \qquad , \qquad  \widetilde{Z}_\text{bulk} \equiv \frac{Z_\text{bulk}}{ Z^\text{Rindler}_\text{bulk}(\beta_H)}  \; .
\end{align}
Here $Z_\text{bulk}=\Tr e^{-\beta_H \hat H}$ is formally defined by \eqref{intro:log Z bulk}, while $Z^\text{Rindler}_\text{bulk}(\beta_H)$ is analogously defined but on a {\it Rindler wedge} of inverse temperature $\beta_H$.\footnote{By this we mean the wedge described by the metric $ds^2 = e^{\frac{4\pi}{\beta_H}x} \left( -dt+dx^2 \right)+r_H^2 d\Omega_{d-1}^2$, which is natural for an observer (at $x=0$) with proper time $t$ and proper acceleration $2\pi/\beta_H$.} The ratio $Z_\text{bulk}/Z^\text{Rindler}_\text{bulk}(\beta_H)$ is understood in a limiting sense explained in Section \ref{sec:char DOS}. Back to the question in the beginning, \eqref{introeq:PI} suggests that the 1-loop Euclidean path integral is in fact computing a relative or renormalized partition function. As we will discuss in Section \ref{sec:canfn}, such a renormalized partition function has an intuitive physical interpretation from the perspective of a near-horizon observer. While we have formally established \eqref{introeq:PI}, quantities in this relation are UV-divergent and require regularization; in the framework of low-energy effective theory of gravity plus matter, these divergences are absorbed into the renormalization of the cosmological constant, Newton's constant and couplings to higher curvatures \cite{Susskind:1994sm,Demers:1995dq}.

From the algebraic QFT point of view (see \cite{Witten:2018zxz} for a recent review), the infinity of $\rho(\omega)$ is related to the fact that the algebra of observables for the scalar QFT outside the horizon is a von Neumann algebra of Type III, which does not admit a trace. It was pointed out recently that including 1-loop effects of gravity turns the algebra from Type III to Type II, for which a trace can be defined up to an arbitrary renormalization \cite{Witten:2021unn,Chandrasekaran:2022cip}. We will comment more on this with some suggestive observations as we conclude in Section \ref{sec:discussion}.

To explain these ideas, we focus almost exclusively on the case of a massive scalar in this work. In  an upcoming paper \cite{BHedge}, we extend our discussions to arbitrary spinning fields, for which other subtleties and qualitatively new features would appear, as already noted in the context of de Sitter space \cite{Anninos:2020hfj,Law:2020cpj}.

\paragraph{Plan of the paper}

In Section \ref{sec:DHS}, we review the DHS formula and introduce the QNM character. In Section \ref{sec:char DOS} we explain the physics of the QNM character by recasting the free scalar theory into a 1D scattering problem, after which we are naturally led to the proposal \eqref{introeq:PI}. In Section \ref{sec:BTZ} and \ref{sec:narscal} we support \eqref{introeq:PI} by the examples of scalars on BTZ and Nariai. In Section \ref{sec:flat} we comment on the case of black holes in asymptotically flat space. We conclude with some remarks in Section \ref{sec:discussion}.

\section{Comments on the Denef-Hartnoll-Sachdev formula}\label{sec:DHS}

In this section we review the DHS formula \cite{Denef:2009kn}, after which we introduce the ``QNM character". For our purpose of getting
the formula \eqref{eq:DHSintegral} as soon as possible, we proceed formally neglecting UV-divergences. A more rigorous treatment is postponed until we discuss explicit examples in Section \ref{sec:BTZ} and \ref{sec:narscal}. The following discussion applies to ($d+1$)-dimensional static spherically symmetric backgrounds ($d\geq 1$):
\begin{align}\label{sym metric}
	ds^2=-F(r)\,dt^2+\frac{dr^2}{F(r)}+r^2 d\Omega_{d-1}^2\, .
\end{align}
Here $d\Omega_{d-1}^2$ is the metric on the unit $S^{d-1}$. There is a horizon at $r=r_H$ if $F(r_H)=0$, with an associated Hawking temperature 
\begin{align}
	\beta_H =\frac{1}{T_H}=\frac{4\pi}{F'(r_H)}\; .
\end{align}
We restrict ourselves to the case where $T_H$ is non-zero. Wick-rotating $t=-i t_E$ and identifying $t_E \simeq t_E +\beta_H$ in \eqref{sym metric}, we obtain a smooth geometry,
\begin{align}\label{E sym metric}
	ds^2 \to ds_E^2=F(r)\, dt_E^2+\frac{dr^2}{F(r)}+r^2 d\Omega_{d-1}^2 \; ,
\end{align}
that arises as a saddle point in the Euclidean gravitational path integral. The horizon at $r=r_H$ is mapped to the origin, near which we can make a change of variables
\begin{align}
	\rho^2 =\frac{4}{F'(r_H)}(r-r_H)\; ,\quad \theta = \frac{2\pi}{\beta_H}t_E
\end{align}
so that the space takes the product form
\begin{align}\label{near hor metric}
	ds^2\approx d\rho^2 +\rho^2d\theta^2+r_H^2\, d\Omega_{d-1}^2\, .
\end{align}

The 1-loop corrections to the gravitational path integral are given by integrating out quadratic fluctuations of matter fields (including the graviton) living on \eqref{E sym metric}. For a real scalar $\phi$ with mass $m^2$, this is given by a functional determinant of a Laplace operator
\begin{align}\label{scalar PI}
	Z_\text{PI}(m^2)=\int\mathcal{D}\phi \, e^{-\frac{1}{2}\int \left(\nabla \phi \right)^2 + m^2 \phi^2}=\frac{1}{\det \left(-\nabla^2+m^2 \right)^{1/2}} \; .
\end{align}
We demand the functions in the functional integration to be regular at the origin $\rho=0$. In terms of the complex coordinate $u=\rho \, e^{-i\theta}$, this means that $\phi$ has a Taylor expansion in $u$ and $\bar{u}$. More precisely, a mode with thermal frequency $k$ has the following $\rho\to 0$ behavior
\begin{align}\label{quan asym}
	\phi_k \sim \rho^{|k|} e^{-ik\theta} =
	\begin{cases}
		u^k \; ,\quad &k\geq 0\\
		\bar{u}^{-k} \; ,\quad &k\leq 0
	\end{cases}\; .
\end{align}
As part of the definition of the path integral, $\phi$ is typically required to satisfy other boundary conditions (e.g. standard or alternate boundary condition in asymptotically $AdS$ black holes).


The key result in \cite{Denef:2009kn} is that 
\begin{align}\label{eq:qnm det formula}
	D(m^2)\equiv \frac{1}{\det \left(-\nabla^2+m^2 \right)}
	=&\prod_{z,\bar{z}}\prod_{k=-\infty}^\infty \left( |k|+ \frac{iz}{2\pi T_H}\right)^{-N_z/2}\left( |k|- \frac{i\bar{z}}{2\pi T_H}\right)^{-N_{\bar z}/2} \; .
\end{align}
Here $z=z(m^2)$ and $\bar z=\bar z(m^2)$ are the QNM and anti-QNM frequencies in the Lorentzian signature, with degeneracies $N_z$ and $N_{\bar z}$ respectively. 

The idea of deriving \eqref{eq:qnm det formula} is the following. We assume that the function $D(m^2)$ is a meromorphic function on the complex $m^2$-plane, and try to match its poles and zeros.\footnote{\label{fn:analytic} This is a very strong assumption; in fact, from the scattering point of view discussed in the next section, there is a natural proposal for how this should be modified.} The observation is that whenever we vary $m^2$ such that $ \frac{iz(m^2)}{2\pi T_H}=-|k|$ or $ \frac{i\bar z(m^2)}{2\pi T_H}=|k|$, the Lorentzian mode with frequency $z$ or $\bar z$ Wick-rotates to a regular mode in the Euclidean signature while it solves the equation of motion $\left(-\nabla^2+m^2\right)\phi =0$, and thus hitting a pole of $D(m^2)$. Since $D(m^2)$ has no zeros, it has the same analytic structure as the function \eqref{eq:qnm det formula}. This completes the argument.\footnote{Generally there is a holomorphic function $e^{P(m^2)}$ multiplying \eqref{eq:qnm det formula}, which is related to its UV-divergence and can be determined by comparing $m^2\to \infty$ asymptotics of \eqref{eq:qnm det formula} and the heat kernel coefficients \cite{Denef:2009kn}.} A similar reasoning for a Dirac spinor leads to \cite{Denef:2009kn}
\begin{align}\label{eq:qnm fer formula}
	Z_\text{PI}(m^2)=\det\left(\slashed{\nabla}+m\right)=\prod_{z, \bar z}\prod_{k=0}^\infty \left( |k|+\frac{1}{2}+ \frac{iz}{2\pi T_H}\right)^{N_z/2}\left( |k|+\frac{1}{2}- \frac{i\bar z}{2\pi T_H}\right)^{N_{\bar z}/2} \; .
\end{align}

When the theory is PT-symmetric, $\bar z$ can be taken to be the complex conjugate of $z$. Alternatively, we observe that for a QNM with frequency $z$, there is an anti-QNM with frequency $-z$. Therefore, we can replace $\bar z \to -z$ in \eqref{eq:qnm det formula} and \eqref{eq:qnm fer formula}, and we have simply
\begin{align}\label{eq:qnmPI}
    Z_\text{PI}(m^2)= \prod_z \prod_{k=-\infty}^\infty \left( |k|+ \frac{iz}{2\pi T_H}\right)^{-N_z/2}
\end{align}
for a scalar and 
\begin{align}\label{eq:qnmPIfer}
	Z_\text{PI}(m^2)=\prod_{z}\prod_{k=0}^\infty \left( |k|+\frac{1}{2}+ \frac{iz}{2\pi T_H}\right)^{N_z} 
\end{align}
for a Dirac spinor. We will focus on this case from now on.

\subsection{The quasinormal mode character}\label{sec:scalar char}

Using $\log x = -\int_0^\infty \frac{dt}{t} e^{-x t}$ (ignoring the issue of UV-divergence), we can formally write \eqref{eq:qnmPI} as
\begin{align}\label{eq:DHSchar}
\log Z_\text{PI} = \int_0^\infty \frac{dt}{2t}\sum_{z}  \sum_{k=-\infty}^\infty N_z \, e^{-\left( |k|+ \frac{iz}{2\pi T_H}\right)t} =\int_0^\infty \frac{dt}{2t}\frac{1+e^{-2\pi t/\beta_H}}{1-e^{-2\pi t/\beta_H}}\,\chi_\text{QNM}(t)\; .
\end{align}
In the second equality we performed the sum over $k$ and scaled $t\to 2\pi t/\beta_H$. Here we have defined a ``QNM character"
\begin{align}\label{eq:qnm char}
\chi_\text{QNM}(t)\equiv \sum_{z} \, N_z  \, e^{-izt} \; .
\end{align}
The formula \eqref{eq:qnmPIfer} for a Dirac spinor can be similarly manipulated. To summarize, the DHS formula for scalars and spinors in an arbitrary static background takes the form
\begin{align}\label{eq:DHSintegral}
\log Z_\text{PI} =\int_0^\infty \frac{dt}{2t}\left( \frac{1+e^{-2\pi t/\beta_H}}{1-e^{-2\pi t/\beta_H}}\chi_\text{QNM}^\text{scalar}(t)-\frac{2\, e^{-\pi t/\beta_H}}{1-e^{-2\pi t/\beta_H}}\chi_\text{QNM}^\text{spinor}(t) \right).
\end{align}
Note that the sums over $k\in \mathbb{Z}$ give the integration kernels capturing bosonic or fermionic statistics. Formula \eqref{eq:DHSintegral} takes the same form as \eqref{introeq:DHSintegral} derived in the context of de Sitter space \cite{Anninos:2020hfj}, where the QNM character $\chi(t)$ is the Harish-Chandra character for a unitary irreducible representation of the de Sitter group. For a generic black hole, the QNM character \eqref{eq:qnm char} might not have such a group theoretic interpretation; however, we will abuse the terminology and call it a ``character". 

Integrals like \eqref{eq:DHSintegral} are UV-divergent at $t\approx 0$ and require regularization, such as heat kernel \cite{Vassilevich:2003xt} or zeta function \cite{Hawking:1976ja} regularization. We will comment more on this when we discuss explicit examples in Section \ref{sec:BTZ} and \ref{sec:narscal}.


\section{Black hole scattering and partition functions}\label{sec:char DOS}

In this section we explain the physics of the QNM character and thus formula \eqref{eq:DHSintegral} from a purely Lorentzian point of view. To that end we will recast our problem into that of 1D scattering; the subsequent discussion is largely inspired by what is known as the Krein–Friedel–Lloyd formula in condensed matter literature. For concreteness we focus on the case of asymptotically AdS black holes, but our discussion is easily generalized to cases with zero or positive cosmological constants.


 
\subsection{Scattering phase shifts and density of states}\label{sec:scattering}

\begin{figure}[H]
\centering
  \includegraphics[scale=0.425]{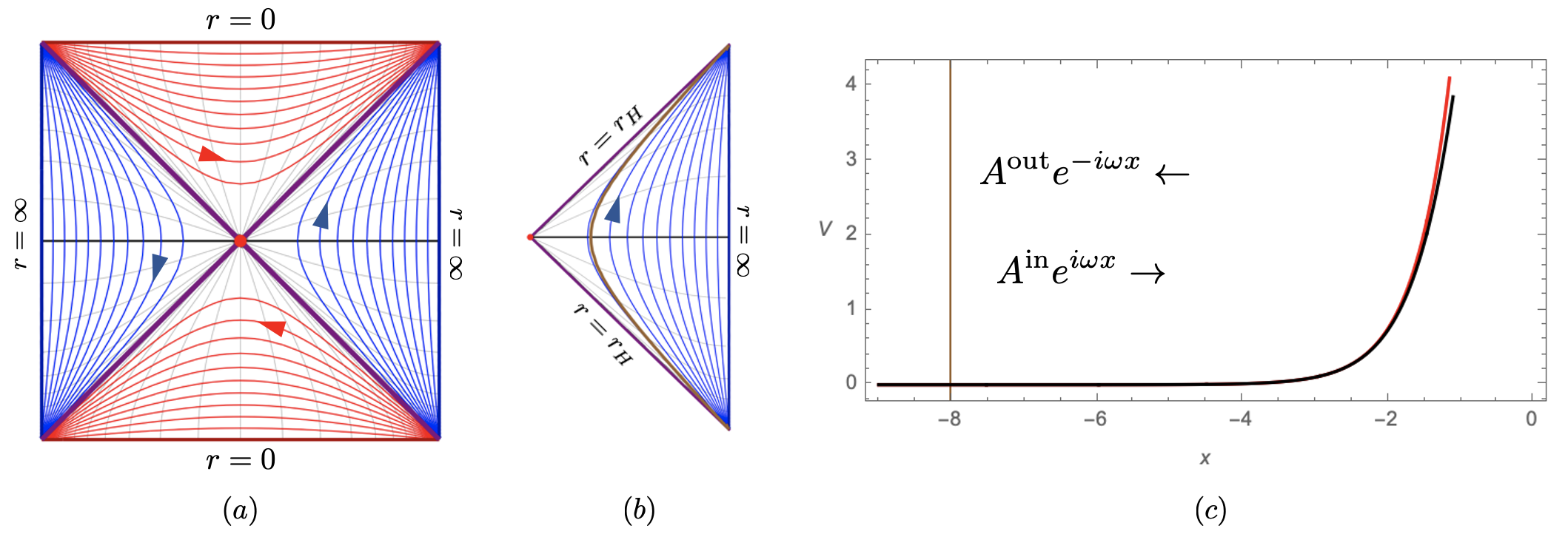}
 \caption{(a) The Penrose diagram for the full two-sided static BTZ black hole geometry. The arrows point in the direction of increasing $t$. (b) The Penrose diagram for the 1-sided geometry, where the $t$-translation generator is interpreted as the Hamiltonian. (c) We plot the scattering potential \eqref{eq:potential} (black) for a scalar with $\Delta=2.1$ and $|l|=3$ on a static BTZ black hole with $r_H=1$, which is hardly distinguishable from the Rindler potential (red) for $x\ll 0$. The brown lines in (a) and (b) indicate the brick wall regulator \cite{tHooft:1984kcu} at $x=-R$, which we will remove at the end. In these figures $R=8$, corresponding to $r\approx 1.13$.}
 \label{fig:penrose scatter}
\end{figure}

To start with, we separate
\begin{align}\label{eq:normalmodes}
	\phi_{\omega l} (t,r,\Omega)= e^{-i \omega t}\,\frac{\psi_l (r)}{r^\frac{d-1}{2} } \, Y_l(\Omega)\; . 
\end{align}
Here $Y_l$ are the $(d-1)$-dimensional spherical harmonics with degeneracy $D_l^d=\frac{2l+d-2}{d-2}\binom{l+d-3}{d-3}$ satisfying $-\nabla^2_{S^{d-1}}Y_l=l(l+d-2)Y_l$. With \eqref{eq:normalmodes} the Klein-Gordon equation $\left( -\nabla^2+m^2\right)\phi=0 $ on the background \eqref{sym metric} is cast into a 1D Schr\"odinger form for each $SO(d)$ quantum number $l\geq 0$:\footnote{This is for $d\geq 3$. The $d=1$ case is trivial; when $d=2$, the $SO(2)\simeq U(1)$ harmonics are $\frac{1}{\sqrt{2\pi}}e^{-il\theta}$ and $l\in \mathbb{Z}$.}
\begin{align}\label{eq:eff Sch}
	\left( -\partial_{x}^2 +V_l(x) \right) \psi_l (x) =\omega^2 \psi_l (x)\; ,
\end{align}
with the effective potential 
\begin{align}\label{eq:potential}
	V_l(x) = F(r) \left[ \frac{d-1}{2r^{\frac{d-1}{2}}} \partial_r \left(r^{\frac{d-3}{2}} F(r) \right) +\left( \frac{l(l+d-2)}{r^2}+m^2\right)  \right] \, .
\end{align}
Here we have introduced the tortoise coordinate
\begin{align}\label{tortoise}
	x \equiv \int^r_\infty\frac{dr'}{F(r')} \; .
\end{align}
We have $x \to -\infty$ as $r\to r_H$, and the integration constant is chosen such that $x\to 0$ as $r\to \infty$. For example, a static BTZ black hole with mass $M$ has $F(r) = \frac{r^2-r_H^2}{\ell_\text{AdS}^2}$ where $\ell_\text{AdS}$ is the AdS length and $r_H\equiv M \ell_\text{AdS}^2$, and we have $x = \frac{\ell_\text{AdS}^2}{2r_H} \log \left( \frac{r-r_H}{r+r_H}\right)$. See Figure \ref{fig:penrose scatter}. We set $\ell_\text{AdS}=1$ from now on. 

\paragraph{Scattering states}

For asymptotically AdS black holes, the general solution to \eqref{eq:eff Sch} is a linear combination of a normalizable and a non-normalizable mode at spatial infinity: 
\begin{align}
    \psi_l (x\to 0 ) \sim C_l^\text{n.} (-x)^\Delta +C_l^\text{n.n.} (-x)^{d-\Delta} \; , \qquad \Delta\equiv \frac{d}{2}+ \sqrt{\frac{d^2}{4}+m^2} \; .
\end{align}
We impose standard quantization where we set $C_l^\text{n.n.}=0$. This picks out the unique solution with no wave transmitted to spatial infinity. This reflects the intuition that the negative cosmological constant creates an infinite gravitational well at spatial infinity ($r\to \infty$), from which any wave must bounce back. Near the horizon, we have a mixture of incoming and outgoing waves:
\begin{align}\label{eq:nearhor}
	\psi_l (x\to -\infty ) \sim A_l^\text{out}(\omega)\, e^{-i\omega x} + A_l^\text{in}(\omega) \, e^{i\omega x}  \, .
\end{align}
Here by ``in" (``out") we mean the waves travel away from (towards) the horizon, as opposed to the common terminology in studies of QNMs. See Figure \ref{fig:penrose scatter}. Since \eqref{eq:eff Sch} is invariant under $\omega\to-\omega$, we have $A_l^{\text{in}} (\omega)=A_l^{\text{out}} (-\omega)$. For real $\omega$, $A_l^\text{in}(\omega)={A_l^\text{out}}^*(\omega)$, and the ratio
\begin{align}\label{eq:s matrix cos}
	\mathcal{S}_l (\omega) = \frac{A_l^\text{out}(\omega) }{A_l^\text{in}(\omega) } \equiv e^{2i\theta_l(\omega)}
\end{align}
is a pure phase, or a rank-1 unitary S-matrix. In terms of the phase shift $\theta_l(\omega)$, \eqref{eq:nearhor} becomes
\begin{align}
	\psi_l (x\to -\infty ) \sim e^{-i\left( \omega x-\theta_l\right)} + e^{i\left( \omega x-\theta_l\right)} \propto \cos \left( \omega x-\theta_l\right)  \; .
\end{align}

\paragraph{Single-particle density of states}

\begin{figure}[ht]
\centering
  \includegraphics[scale=0.4]{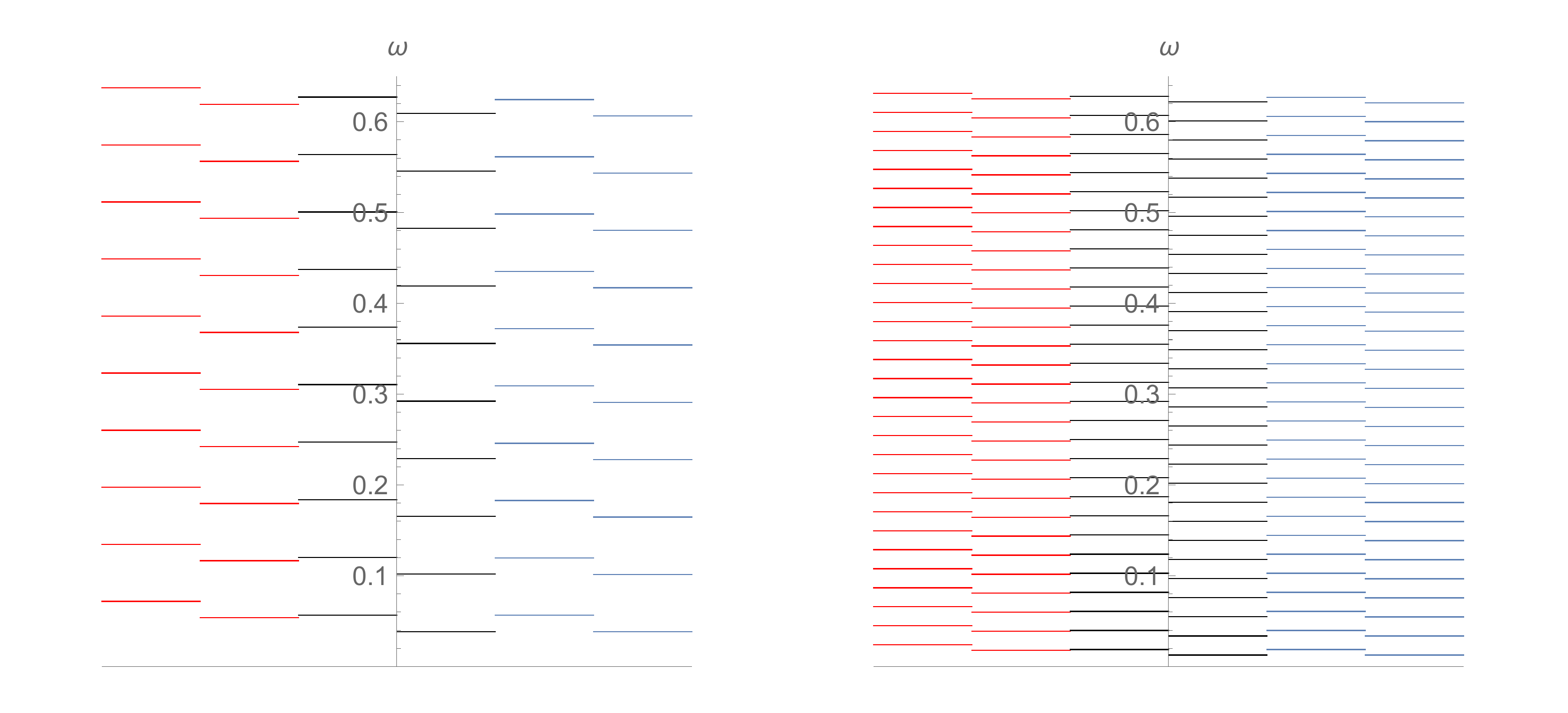}
 \caption{Left: The black lines mark the discrete spectrum obtained by numerically solving \eqref{eq:disentower} for a scalar with $\Delta=2.1$ and $|l|=3$ on a static BTZ black hole with $r_H=1$. We compare this with the cases of flat ($\bar V(x)=0$) (red) and Rindler ($\bar V_l(x)= 4 \,e^{2 x}$) (blue) potential. For each case we have arranged the $\pm$-towers \eqref{eq:disentower} on the left (right). For the left and right figures we have chosen $R=100$ and $R=300$ respectively. Observe that the energy level spacing scales roughly as $\frac{2\pi}{R}$, and we will have a continuum of energies as $R\to\infty$. In both figures we choose $\xi_0=0.9$, which only shifts the relative heights of the $\pm$-towers but not the spacing within the individual tower.}
 \label{fig:spectrum}
\end{figure}

For every $\omega>0$, there is a unique solution to \eqref{eq:eff Sch} subject to the standard boundary condition. In other words, within any interval $\Delta \omega$, the number $\rho_l(\omega) \Delta \omega$ of normalizable solutions is infinite, implying in particular that the single-particle density of states (DOS) $\rho_l(\omega)$ is infinite. More explicitly, we can cut off the scattering problem \eqref{eq:eff Sch} at a large distance $x=-R$, and impose a Dirichlet-type boundary condition, i.e. set $\cos \left(\omega R+\theta_l (\omega)\right) = \psi_0$ for some constant $\psi_0$. See (b) and (c) in Figure \ref{fig:penrose scatter}. This implies the quantization condition
\begin{align}\label{eq:disentower}
    \omega_n R+\theta_l (\omega_n)= 2n \pi \pm \xi_0 \; , \qquad \xi_0 \equiv \cos^{-1} \psi_0 \; , \qquad n=0,1,2, \cdots .
\end{align}
In Figure \ref{fig:spectrum}, we obtain the discrete spectra for various potentials by directly solving \eqref{eq:disentower}. For large $R$, the spacing $\Delta\omega_n = \omega_{n+1}-\omega_n$ between consecutive levels become small, and we have $\theta_l(\omega_{n+1}) \approx \theta_l(\omega_n)+ \Delta \omega_n \theta_l'(\omega_n)$. We can then compute the smoothed-out DOS 
\begin{align}\label{eq:regularDOS}
    \rho_l^R (\omega) \equiv \frac{1}{\Delta \omega}= \frac{R + \theta_l'(\omega)}{\pi}+O\left(\frac{1}{R} \right) \; .
\end{align}
We took into account contributions from both $\pm$ towers \eqref{eq:disentower}. 

So far, working with a finite cutoff at $x=-R$ is no different from the brick wall model \cite{tHooft:1984kcu}. Because of the leading ``brick wall term'' $\frac{R}{\pi}$, \eqref{eq:regularDOS} diverges as $R\to \infty$. \footnote{This leading behavior is an example of what is known as the Weyl Law in spectral theory.} However, notice that this universal $\frac{R}{\pi}$ term  contains {\it no information at all} about the potential $V_l(x)$: It is present for any system in a box with width $R$ large compared to the range of $V_l(x)$. In contrast, the subleading term $\frac{\theta_l'(\omega)}{\pi}$ does depend on the shape of $V_l(x)$. In the context of black hole physics, $V_l(x)$ contains information about the black hole geometry \eqref{sym metric}, as well as the mass $m^2$ and angular momentum $l$ of the matter field.

\begin{figure}[ht]
\centering
  \includegraphics[scale=0.45]{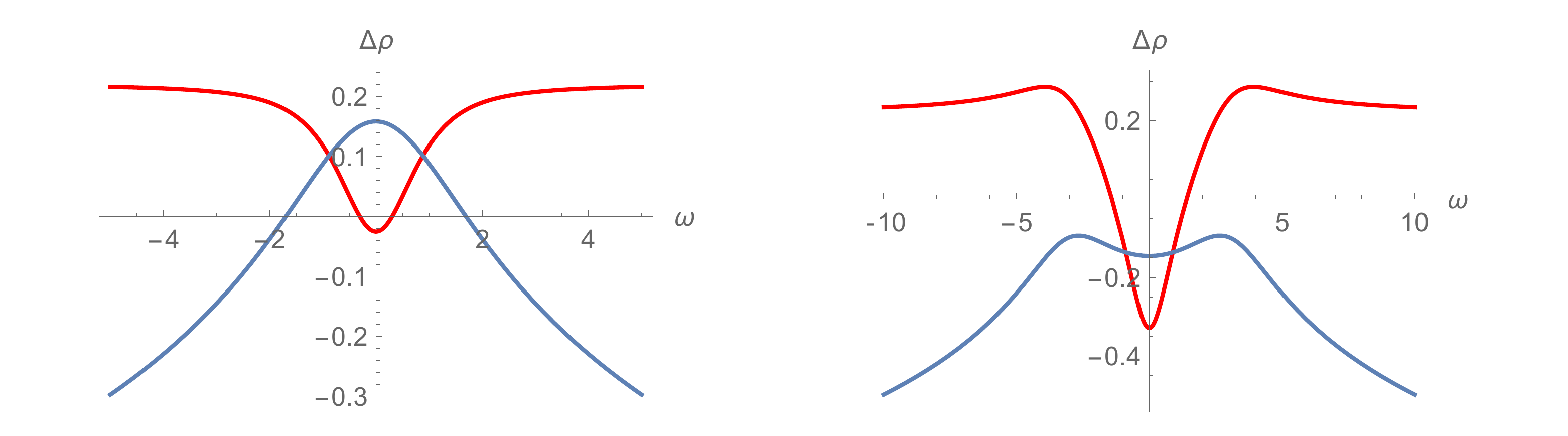}
 \caption{Plots of $\Delta \rho_l(\omega)$ for $\omega \in \mathbb{R}$ for a scalar with $\Delta=2.1$ on a static BTZ black hole with $r_H=1$ when $l=0$ (left) and $|l|=3$ (right). The red and blue line correspond to the minimal (with $\bar V_l(x)=0$) and the Rindler (with $\bar V_l(x)= 4\, e^{2 x}$) references respectively.}
 \label{dosreal}
\end{figure}

A common strategy of sending the cutoff $R$ to infinity while extracting the interesting information about the spectrum encoded in the scattering phase $\theta_l(\omega)$ is the following. We consider a reference problem with potential $\bar V_l(x)$ which we assume falls off fast enough such that the solution $\bar \psi_l(x)$ is still a linear combination of plane waves $e^{\pm i \omega x}$ near $x=-R$. Reasoning as above, we obtain another regulated DOS:
\begin{align}\label{eq:refDOS}
    \bar \rho_l^R (\omega)= \frac{R + \bar\theta_l'(\omega)}{\pi}+O\left(\frac{1}{R} \right) \; .
\end{align}
The key point is that the {\it difference} $\rho_l^R(\omega)-\bar\rho_l^R(\omega)$ remains finite as we take $R\to \infty$ and is proportional to the derivative of the difference of the phase shifts 
\begin{align}\label{friedel relation cos}
	\Delta \rho_l(\omega) \equiv \lim_{R\to \infty}\left(\rho_l^R(\omega)-\bar\rho_l^R(\omega)\right) = \frac{1}{\pi} \partial_\omega \Delta\theta_l (\omega) \; , \qquad \Delta\theta_l (\omega) \equiv \theta_l (\omega)-\bar\theta_l (\omega) \; .
\end{align}
There is still a question of what reference potential $\bar V(x)$ we should choose; in fact, without any other input, there is no canonical choice of $\bar V(x)$. See Figure \ref{dosreal} for an example. Quite amazingly, the Euclidean path integral uniquely picks out a natural one. We turn to this in the next section, before which two more comments are in order. 

First, the relation \eqref{friedel relation cos} does not depend on the choice of boundary conditions at $x=-R$ in either regulated problems; one can even relax the boundary conditions by allowing $\omega$-dependent values $\psi_0(\omega)$ at the cutoff $x=-R$, as long as we take them to be the same for both problems. It is also clear from \eqref{friedel relation cos} that the right hand side only depends on the {\it asymptotic form} of the solutions as $x\to -\infty$ but not their values at $x=-R$.


Second, we have focused on a fixed angular momentum $l$. At each $l\geq 0$, $\Delta \rho_l(\omega)$ is finite. For the full theory we need to sum over all $l\geq 0$:
\begin{align}\label{eq:fullrhochange}
	\Delta \rho(\omega) = \sum_{l=0}^\infty D_l^d \, \Delta \rho_l(\omega)=
	\frac{1}{2\pi i}\partial_\omega \sum_{l=0}^\infty D_l^d \left(\log \mathcal{S}_l(\omega) - \log \bar{\mathcal{S}}_l(\omega)\right) \; .
\end{align}
We have expressed the last equality in terms of the S-matrices $\mathcal{S}_l(\omega)$ and $\bar{\mathcal{S}}_l(\omega)$. As usual in QFT, this sum is typically UV-divergent and requires regularization.

\subsection{Black hole, Rindler, and Euclidean partition functions}\label{sec:canfn}

We consider the thermal canonical partition function for the free scalar living on \eqref{sym metric}
at the black hole temperature
\begin{align}\label{eq:Zbulk}
	Z_\text{bulk} \equiv \Tr \, e^{-\beta_H \hat H} \; .
\end{align}
We use the label ``bulk'' since the ideal thermal gas comprises excitations of scalar quanta in the bulk of the spacetime. $\hat H$ is the Hamiltonian generating $t$ translations. The trace Tr is formally tracing over the Fock space constructed by acting with the creation operators associated with the normal modes \eqref{eq:normalmodes}. Following the standard procedure of canonical quantization and summing over bosonic occupation numbers in the trace \eqref{eq:Zbulk}, we have
\begin{align}\label{log Z bulk}
	\log Z_\text{bulk} = -\int_0^\infty d\omega \, \rho(\omega) \log \left( e^{\beta_H \omega/2}- e^{-\beta_H \omega /2}\right)  \; .
\end{align}
There are two kinds of divergences in this expression. We have the usual UV-divergences coming from summing over all angular momenta $l\geq 0$ and integrating over all energies $\omega>0$.  Within the framework of low-energy effective theory of gravity plus matter, these divergences can be absorbed into the cosmological constant, Newton's constant and local couplings to higher curvature terms in the gravity sector \cite{Susskind:1994sm,Demers:1995dq}.





The second type of divergence is that of the single-particle DOS $\rho(\omega)=\sum_{l\geq 0}\rho_l(\omega)$  originating from the fact that the normal mode spectrum is continuous, for which we extensively discussed in Section \ref{sec:scattering}. We do not commit to interpreting this divergence as either UV or IR; in fact, at each angular momentum $l\geq 0$, $\rho_l(\omega)$ is infinite for {\it all} energies $\omega>0$. Through the procedure described in Section \ref{sec:scattering}, for each $l\geq 0$ we can obtain a finite difference $\Delta\rho_l(\omega)$ (understood in the limiting sense \eqref{friedel relation cos}) between $\rho_l(\omega)$ and $\bar\rho_l(\omega)$, the DOS for a reference system.


\paragraph{Euclidean path integral and the renormalized canonical partition function}

Inspired by the discussion in Section \ref{sec:scattering}, instead of \eqref{log Z bulk} we are led to consider a quantity
\begin{align}\label{eq:renZbulk}
    \log Z_\text{bulk} - \log \bar Z_\text{bulk}= -\int_0^\infty d\omega \, \Delta\rho(\omega) \log \left( e^{\beta_H \omega/2}- e^{-\beta_H \omega /2}\right)  \; .
\end{align}
Here $\bar Z_\text{bulk}$ is a thermal canonical partition function for a system with a reference DOS $\bar\rho(\omega)$. This proposal of considering a difference $\log Z_\text{bulk} - \log \bar Z_\text{bulk}$ instead of $\log Z_\text{bulk}$ itself  is similar to that of considering relative entropy rather than entanglement entropy \cite{Jafferis:2015del}. A priori {\it any} reference $\bar Z_\text{bulk}$ (for example one for a system with a strictly flat potential $\bar V=0$) would lead to a finite difference $\Delta \rho_l(\omega)$ for each $l\geq 0$. Therefore, we have a ``renormalized" partition function $Z_\text{bulk}/\bar Z_\text{bulk}$, understood in the limiting sense \eqref{friedel relation cos}, defined up to an arbitrary choice of $\bar Z_\text{bulk}$. 


What fixes an answer is the Euclidean path integral \eqref{scalar PI}. Our key observation is that choosing $\bar Z_\text{bulk}$ to be the canonical partition function for the free scalar living on the Rindler-like wedge
\begin{align}\label{eq:refM}
     ds^2 = e^{\frac{4\pi}{\beta_H}x} \left( -dt^2+dx^2 \right)+r_H^2 d\Omega_{d-1}^2 \; , \qquad -\infty<x<\infty \; ,
\end{align}
the quantity \eqref{eq:renZbulk} is exactly equal to $\log Z_\text{PI}$. In other words, we claim that
\begin{align}\label{eq:ratioPI}
    \widetilde{Z}_\text{bulk} \equiv \frac{Z_\text{bulk}}{Z^\text{Rindler}_\text{bulk}(\beta_H)} = Z_\text{PI} \; , \qquad Z^\text{Rindler}_\text{bulk}\equiv\Tr \,e^{-\beta_H \hat{H}_0}
\end{align}
where $\hat{H}_0$ is the Hamiltonian generating $t$-translation in \eqref{eq:refM} and Tr is formally tracing over the Fock space. We support the relation \eqref{eq:ratioPI} with the examples of BTZ and Nariai in Section \ref{sec:BTZ} and \ref{sec:narscal}; one can also check that it holds for the case of static patch in de Sitter space (noting that our $\widetilde{Z}_\text{bulk}$ is called $Z_\text{bulk}$ in \cite{Anninos:2020hfj}).

\paragraph{The Rindler-like region near horizon} 

The relation \eqref{eq:ratioPI} means that $\log Z_\text{PI}$ has the following Lorentzian interpretation. We start by observing that the region near horizon takes the form of a product of a 2D Rindler space and a $(d-1)$-dimensional sphere with radius $r_H$:
\begin{align}\label{eq:Rindlerlike}
	ds^2 \approx e^{\frac{4\pi}{\beta_H}x} \left( -dt^2+dx^2 \right)+r_H^2 d\Omega_{d-1}^2\quad \text{as} \quad x\to -\infty \; .
\end{align}
The scattering problem \eqref{eq:eff Sch} in this region reduces to
\begin{align}\label{eq:Rindlerscatter}
    \left( -\partial_x^2 + M_l^2 e^{\frac{4\pi}{\beta_H}x} \right) \psi(x) = \omega^2 \psi(x) \; , \qquad M_l \equiv \sqrt{\frac{l(l+d-2)}{r_H^2}+m^2} \; .
\end{align}
Defining $x_l \equiv \frac{ \beta_H }{2 \pi } \log \frac{\beta_H M_l}{4\pi}$, this is equivalent to
\begin{align}\label{eq:nearhorscatt}
    \left[ -\partial_{x'}^2 + V^\text{Rindler} (\beta_H,x') \right] \tilde\psi(x') = \omega^2 \tilde\psi(x') \; , \qquad V^\text{Rindler} (\beta, x)\equiv \left(\frac{4\pi}{\beta} \right)^2 e^{\frac{4\pi}{\beta}x} \; .
\end{align}
where $x' \equiv x+x_l$ and $\tilde\psi(x') \equiv \psi(x'-x_l)$. This Schr\"odinger equation is same as that of the spacelike Liouville quantum mechanics.\footnote{We thank Daniel Kapec for pointing this out.} Notice that the information about the black hole geometry (except for its temperature $T_H$) and the scalar (its mass and angular momentum) becomes completely invisible. A near-horizon observer studying \eqref{eq:nearhorscatt} would not be able to distinguish the black hole spacetime \eqref{sym metric} and the Rindler-like wedge \eqref{eq:refM} (see Figure \ref{fig:penrose scatter}); they would obtain an S-matrix (see Appendix \ref{app:rindler} for details)
\begin{align}\label{eq:RindlerS}
    \mathcal{S}^\text{Rindler} (\beta_H,\omega) = \frac{\Gamma
   \left(\frac{i \beta_H  \omega }{2 \pi }\right)}{\Gamma
   \left(-\frac{i \beta_H  \omega }{2 \pi }\right)} \; .
\end{align}
and the associated regularized DOS. If they probe much further so that they detect the non-trivial features of $V_l(x)$, they would then detect a {\it change} in the DOS $\Delta \rho(\omega)$ and thus the free energies $\log Z_\text{bulk}-\log Z^\text{Rindler}_\text{bulk}(\beta_H)$. The relation \eqref{eq:ratioPI} means that $\log Z_\text{PI}$ measures this change.

\paragraph{Connection with DHS formula}


From \eqref{eq:fullrhochange}, we observe that $\Delta \rho(\omega)$ hits a pole $z$ whenever $A_l^\text{in}(z)=0$. These are precisely the QNM frequencies. Since $A_l^\text{in}(\omega)=A_l^\text{out}(-\omega)$ for any $l$, $\Delta \rho(\omega)$ must hit another pole at the anti-QNM frequency $\omega=-z$. These poles contribute to $\Delta \rho(\omega)$ as 
\begin{align}\label{eq:renDOS}
    \Delta \rho_\text{QNM}(\omega) = \frac{1}{2\pi i} \sum_z N_z \left(\frac{1}{\omega+z}-\frac{1}{\omega-z}\right) =\int_0^\infty \frac{dt}{2\pi} \left(e^{i\omega t} +e^{-i\omega t}  \right) \chi_\text{QNM}(t) 
\end{align}
where in the last equality we have formally written in terms of the QNM character \eqref{eq:qnm char}.  While in principle there could be a holomorphic part contributing to $\Delta \rho(\omega)$, in all the explicit examples we have checked, $\Delta \rho(\omega)$ does not receive such a contribution and $\Delta \rho_\text{QNM}(\omega)$ gives the complete answer. In such case, substituting \eqref{eq:renDOS} into \eqref{eq:renZbulk} and doing the $\omega$-integral,\footnote{We resolve the $t^{-2}$ pole in the factors multiplying $\chi_\text{QNM}(t)$ by $t^{-2} \to \frac{1}{2}\left((t+i \epsilon)^2+(t-i\epsilon)^2 \right)$.} we have
\begin{align}\label{eq:difffreeen}
    \log \widetilde{Z}_\text{bulk}= \int_0^\infty \frac{dt}{2t}\frac{1+e^{-2\pi t/\beta_H}}{1-e^{-2\pi t/\beta_H}}\chi_\text{QNM}(t)\; .
\end{align}
The right hand side is precisely the bosonic part of \eqref{eq:DHSchar}. 

\begin{figure}[ht]
\centering
  \includegraphics[scale=0.5]{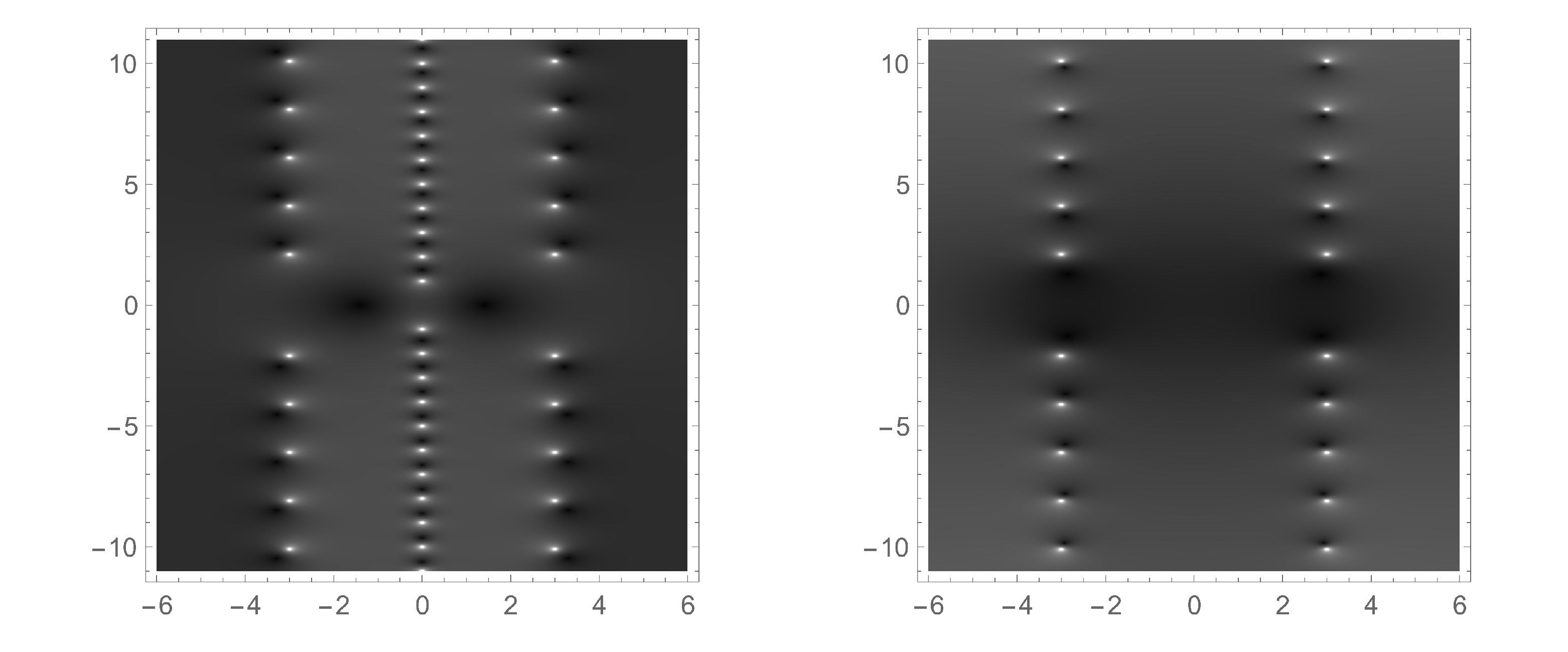}
 \caption{Plots of $|\Delta \rho_{|l|=3}(\omega)|$ on the complex $\omega$-plane for a scalar with $\Delta=2.1$ on a static BTZ black hole with $r_H=1$. Lighter is larger with plot range 0 (black) $<\Delta\rho<4$ (white). When the minimal (with $\bar V_l(x)=0$) reference is chosen (left), there are poles lying on the imaginary axis; these poles are completely absent when the Rindler (with $\bar V_l(x)= 4\, e^{2x}$) reference is taken instead (right). The common poles for these two plots are the QNM frequencies \eqref{BTZ qnm scalar}.}
 \label{doscomplex}
\end{figure}

Generically, $A_\text{out/in}(\omega)$ have poles, which potentially leads to more poles of $\Delta \rho(\omega)$. The statement \eqref{eq:ratioPI} predicts that these would-be poles are canceled by the Rindler S-matrix \eqref{eq:RindlerS}. We confirm this and thus \eqref{eq:ratioPI} for scalars on static BTZ (Section \ref{sec:BTZ}), Nariai (Section \ref{sec:narscal})  and de Sitter static patch (Appendix \ref{app:dS}). In Figure \ref{doscomplex}, with an example of a scalar on static BTZ, we show a comparison of $\Delta\rho_l(\omega)$ on the complex $\omega$-plane for two different choices of reference.

While we have focused on the case of scalars, it is straightforward to generalize our arguments to a Dirac spinor. Instead of \eqref{eq:renZbulk} we have
\begin{align}\label{eq:ferrenZbulk}
	\log Z_\text{bulk}  - \log \bar{Z}_\text{bulk}= \int_0^\infty d\omega \, \Delta\rho(\omega) \log \left( e^{\beta_H \omega/2}+ e^{-\beta_H \omega /2}\right) \; .
\end{align}
Substituting \eqref{eq:renDOS} into this gives the fermionic part of \eqref{eq:DHSchar}.

\paragraph{Retarded Green functions}

Instead of S-matrices and phase shifts, we could have phrased our discussions in terms of retarded Green functions for the scattering problems:
\begin{align}
    \Delta \rho (\omega)  = - \frac{1}{\pi} \Im \left( G(\omega+i\epsilon)-\bar G(\omega+i\epsilon) \right) \; ,
\end{align}
which could provide a link between this work with previous studies in Lorentzian AdS/CFT \cite{Son:2002sd} or the de Sitter static patch \cite{Anninos:2011af}, for example. If the effective potential \eqref{eq:eff Sch} falls off exponentially, the retarded Green function and thus $\Delta \rho (\omega)$ only has poles on the complex $\omega$-plane \cite{Ching:1994bd,Ching:1995tj}. The potentials for the exactly computable examples of BTZ, Nariai, static patch of de Sitter are all of the P\"{o}schl-Teller type and thus satisfy this condition.

For generic black holes, $\Delta \rho (\omega)$ could have more complicated analytic structures on the complex $\omega$-plane such as branch cuts (which in the case of asymptotically flat black holes lead to the so-called Price's tail \cite{PhysRevD.5.2419}, a long-time power-law fall-off of the retarded Green function). Nonetheless, a black hole character can be {\it defined} as the Fourier transform (up to regularization of UV-divergences)
\begin{align}\label{eq:BHchar}
    \chi_\text{BH}(t) \equiv \int_{-\infty}^\infty d\omega \, e^{i\omega t} \Delta \rho (\omega) \; 
\end{align}
Provided we have PT-symmetry so that $\chi_\text{BH}(t)=\chi_\text{BH}(-t)$, \eqref{eq:difffreeen} is generalized to 
\begin{align}\label{eq:difffreeengen}
    \log \widetilde{Z}_\text{bulk}= \int_0^\infty \frac{dt}{2t}\frac{1+e^{-2\pi t/\beta_H}}{1-e^{-2\pi t/\beta_H}}\chi_\text{BH}(t)\; .
\end{align}
In principle one could write down a spectral representation for the black hole character \eqref{eq:BHchar} through deforming the integration contour \eqref{eq:BHchar} on the complex $\omega$-plane, so that the contributions from the different singularity structures can be separated.

Going back to the comment in footnote \ref{fn:analytic}, DHS made a strong assumption about the analytic structure of $\log Z_\text{PI}$. Granting the equality \eqref{eq:ratioPI}, we expect generally the correct answer for $\log Z_\text{PI}$ would be given by \eqref{eq:difffreeengen}. It would be interesting to check this.

\section{Example: Scalar on static BTZ}\label{sec:BTZ}

As a first demonstration, we consider a scalar with mass $m^2 = \Delta (\Delta-2)$ living on a static BTZ background (setting $\ell_\text{AdS}=1$):
\begin{align}\label{eq:btzmetric}
    ds^2 = -\left(r^2-r_H^2 \right) dt^2 + \frac{dr^2}{r^2-r_H^2} +r^2 d\phi^2 \quad ,\qquad  r_H\equiv M =2\pi T_H \; .
\end{align}
Since Euclidean BTZ (EBTZ) is related to thermal $AdS_3$ ($TAdS_3$) by a large diffeomorphism, their path integrals are equal upon the modular transformation
\begin{align}\label{mod trans}
	\tau \to -\frac{1}{\tau}\,\qquad \tau=2\pi i T_H \; .
\end{align}
The 1-loop free energy
of a scalar on \eqref{eq:btzmetric} was first computed in \cite{Mann:1996ze}.

\paragraph{Scattering and DOS}

Solving $(-\nabla^2+m^2) \phi =0$ on \eqref{eq:btzmetric} while imposing the standard boundary condition at spatial infinity, one finds the near-horizon behavior \eqref{eq:nearhor} with
\begin{align}
    A_l^\text{in}(\omega) = A_l^\text{out}(-\omega) \propto \frac{\Gamma \left(-\frac{i\omega }{r_H}\right)}{\Gamma \left(-\frac{i  }{2 r_H}\left(\omega 
   +l \right) +\frac{\Delta}{2}\right) \Gamma \left(-\frac{i  }{2 r_H}\left(\omega 
   -l \right) +\frac{\Delta}{2}\right)}
\end{align}
where $l=0,\pm1,\pm2,\dots$ is the $U(1)$ angular momentum along the spatial circle. Therefore we have
\begin{gather}
    \mathcal{S}_l (\omega) \equiv\frac{A_l^\text{out}(\omega)}{A_l^\text{in}(\omega)} = \mathcal{S}_l^\text{BTZ} (\omega) \mathcal{S}^\text{Rindler} \left(\frac{2\pi}{r_H},\omega\right) \label{eq:btzsmatrix}\\
   \mathcal{S}_l^\text{BTZ} (\omega)\equiv \frac{\Gamma \left(-\frac{i  }{2 r_H}\left(\omega +l \right) +\frac{\Delta}{2}\right) \Gamma \left(-\frac{i}{2 r_H}\left(\omega -l \right) +\frac{\Delta}{2}\right)}{\Gamma \left(\frac{i }{2 r_H}\left(\omega 
   -l \right) +\frac{\Delta}{2}\right) \Gamma \left(\frac{i  }{2 r_H}\left(\omega 
   +l \right) +\frac{\Delta}{2}\right)}  \; .
\end{gather}
Here $\mathcal{S}^\text{Rindler} \left(\beta,\omega\right)$ is the Rindler S-matrix \eqref{eq:RindlerS}. Using these we plot $\Delta \rho_{l}(\omega)$ for real $\omega$ in Figure \ref{dosreal} and $|\Delta \rho_{l}(\omega)|$ on the complex $\omega$-plane in Figure \ref{doscomplex}. Now, choosing the reference S-matrix $\bar{\mathcal{S}}$ in \eqref{eq:fullrhochange} to be $\mathcal{S}^\text{Rindler} \left(\frac{2\pi}{r_H},\omega\right)$ for each $l\in\mathbb{Z}$, the renormalized DOS 
\begin{align}\label{eq:rhobtz}
    \Delta \rho(\omega) = \frac{1}{2\pi i} \partial_\omega \log \det \mathcal{S}^\text{BTZ}(\omega) \; , \qquad \det \mathcal{S}^\text{BTZ}(\omega)= \prod_{l\in \mathbb{Z}} \mathcal{S}_l^\text{BTZ}(\omega) 
\end{align}
has poles only at the QNM frequencies for the scalar.

\paragraph{BTZ character and partition functions}

Our discussion in Section \ref{sec:canfn} guarantees that the renormalized partition function $\widetilde{Z}_\text{bulk}^\text{BTZ}$ is equal to the Euclidean path integral $Z^\text{EBTZ}_\text{PI}$, but let us see how it works explicitly. The scalar in question has the QNM spectrum  \cite{Cardoso:2001hn}
\begin{align}\label{BTZ qnm scalar}
	z_{n,l,\pm} =& \pm l-2\pi T_H i (\Delta+2n) \; , 
\end{align}
where $n=0,1,2,\cdots$ is the overtone number and $l\in\mathbb{Z}$ is the $U(1)$ angular momentum quantum number. These are the poles of \eqref{eq:rhobtz}. With \eqref{BTZ qnm scalar} we obtain the character
\begin{align}\label{btz char scalar}
	\chi^\text{BTZ} (t) =\sum_{n,l,\pm} e^{-i z_{n,l,\pm}t} \; =\frac{4\pi e^{-2\pi T_H\Delta t}}{1-e^{-4\pi T_H t}}\sum_{k\in\mathbb{Z}}  \delta(t-2\pi k)\, .
\end{align}
Here the sum of delta functions comes from the sum over $l\in \mathbb{Z}$. Plugging \eqref{btz char scalar} into the character formula with a UV-cutoff at $t=\epsilon$, the integral is localized to a sum
\begin{align}
	\log \widetilde{Z}^\text{BTZ}_\text{bulk} =\sum_{k=1}^\infty \frac{1}{k}\frac{q_{k}^{\Delta}}{(1-q_{k})^2} \; ,\qquad q_{k}\equiv e^{-(2\pi)^2 T_H k}\; .
\end{align}
This agrees exactly with the $TAdS_3$ result \eqref{appeq:TAdS} upon the modular transformation \eqref{mod trans} as expected.

\section{Example: Scalar on Nariai spacetime}\label{sec:narscal}

To illustrate that our considerations extend to more general spacetimes than asymptotically AdS ones, we study in this section a free scalar on the Nariai spacetime, whose metric is \cite{1950SRToh..34..160N}:
\begin{align}\label{eq:narcoords}
	ds^2=-\left(1-y^2 \right)dt^2 +\frac{\ell^2_N }{1-y^2}\;dy^2+r_N^2 d\Omega_{d-1}^2\; , \qquad -1<y<1\; .
\end{align}
Here $\ell_N$ and $r_N$ are related to the dS length $\ell_\text{dS}$ through
\begin{align}
 \ell_N\equiv \frac{\ell_\text{dS}}{\sqrt{d}} \;, \qquad r_N \equiv  \sqrt{\frac{d-2}{d}}\ell_\text{dS} \;, \qquad \ell_\text{dS} \equiv \sqrt{\frac{d(d-1)}{2\Lambda}}  \; .
\end{align}
This geometry is locally $dS_2\times S^{d-1}$, with isometry group $SO(1,2)\times SO(d)$. There are two horizons (cosmological and black hole) at $y =\pm 1$ with the same Hawking temperature $T_N=\frac{1}{2\pi \ell_N}$. Note that this temperature is higher than the temperature $T_\text{dS}=\frac{1}{2\pi \ell_\text{dS}}$ for pure de Sitter. A possible microscopic realization of the Nariai geometry in matrix theory is recently discussed in \cite{Susskind:2021dfc}. 

\paragraph{Scattering and DOS}

Separating $\phi_{\omega l}(t,y,\Omega) = e^{-i \omega t} \psi(y) \, Y_l(\Omega)$, the Klein-Gordon equation  $(-\nabla^2+m^2) \phi =0$ on \eqref{eq:narcoords} in terms of $x=\ell_N\int_0^y \frac{dy'}{1-{y'}^2}=\frac{\ell_N}{2}\log \frac{1+y}{1-y}$ reads
\begin{align}\label{eq:NariaiSch}
    \left(-\partial_x^2 + \frac{m_l^2}{\cosh^2 \frac{x}{\ell_N}}\right)\psi(x) =\omega^2 \psi(x) \; , \qquad -\infty<x<\infty \;.
\end{align}
Therefore, the problem is decomposed into a tower of Kaluza-Klein (KK) modes living on $dS_2$ labeled by the $SO(d)$ angular momenta $l\geq 0$, each with an effective mass
\begin{align}\label{N eff mass}
	m_l^2=m^2+\frac{l(l+d-2)}{r_N^2} 
\end{align}
and an associated conformal dimension $\Delta_l$ on $dS_2$
\begin{equation}\label{eq:delta_l}
    \Delta_l = \frac12 + i\nu_l\; ,\qquad \nu^2_l = m^2_l \ell^2_N-\frac14 \;, \qquad \bar \Delta_l \equiv 1- \Delta_l \; .
\end{equation}
Notice that there are two asymptotic regions ($x\to \pm \infty$) in the scattering problem \eqref{eq:NariaiSch}, as opposed to the asymptotically AdS case discussed in Section \ref{sec:char DOS}. Also, the two linearly independent solutions $\psi^{(1)}$ and $\psi^{(2)}$ to \eqref{eq:NariaiSch} are both regular at the origin $y=0$, and thus perfectly good solutions. Near the horizons, these solutions become
\begin{align}
    \psi^{(1)}(|x|\to \infty) \propto & \frac{\Gamma\left(i \ell_N \omega\right)}{\Gamma\left(\frac{\Delta_l+i \ell_N \omega}{2}\right) \Gamma\left(\frac{\bar\Delta_l+ i \ell_N \omega}{2}\right)} e^{i \omega |x|}+\frac{\Gamma\left(-i \ell_N \omega\right)}{\Gamma\left(\frac{\Delta_l- i \ell_N \omega}{2}\right) \Gamma\left(\frac{\bar\Delta_l- i \ell_N \omega}{2}\right)} e^{-i \omega |x|} \;, \nn\\
    \psi^{(2)}(|x|\to \infty) \propto &\frac{\Gamma\left(i \ell_N \omega\right)}{\Gamma\left(\frac{1+\Delta_l+i \ell_N \omega}{2}\right) \Gamma\left(\frac{1+\bar\Delta_l+ i \ell_N \omega}{2}\right)} e^{i \omega |x|}+\frac{\Gamma\left(-i \ell_N \omega\right)}{\Gamma\left(\frac{1+\Delta_l- i \ell_N \omega}{2}\right) \Gamma\left(\frac{1+\bar\Delta_l- i \ell_N \omega}{2}\right)} e^{-i \omega |x|} \;.
\end{align}
From this we see that these modes have exactly equal amounts of incoming and outgoing fluxes from either horizons. We have a diagonal S-matrix
\begin{gather}
    \mathcal{S}_l (\omega) =\mathcal{S}^\text{N}_l (\omega)\mathcal{S}^\text{Rindler} (2\pi \ell_N,\omega)\qquad , \qquad \mathcal{S}^\text{N}_l (\omega) \equiv
    \begin{pmatrix}
        \mathcal{S}^{(1)}_l (\omega) & 0 \\
        0 &\mathcal{S}^{(2)}_l (\omega)
    \end{pmatrix}
     \label{eq:nariaismatrix}\\
    \mathcal{S}^{(1)}_l (\omega) = \frac{\Gamma\left(\frac{\Delta_l-i \ell_N \omega}{2}\right) \Gamma\left(\frac{\bar\Delta_l- i \ell_N \omega}{2}\right)}{\Gamma\left(\frac{\Delta_l+i \ell_N \omega}{2}\right) \Gamma\left(\frac{\bar\Delta_l+ i \ell_N \omega}{2}\right)} \quad , \quad \mathcal{S}^{(2)}_l (\omega) = \frac{\Gamma\left(\frac{1+\Delta_l-i \ell_N \omega}{2}\right) \Gamma\left(\frac{1+\bar\Delta_l- i \ell_N \omega}{2}\right)}{\Gamma\left(\frac{1+\Delta_l+i \ell_N \omega}{2}\right) \Gamma\left(\frac{1+\bar\Delta_l+ i \ell_N \omega}{2}\right)}  \; .
\end{gather}
Again, choosing the reference S-matrix to be $\mathcal{S}^\text{Rindler} (2\pi \ell_N,\omega)$ for each $l$, the renormalized DOS 
\begin{align}\label{eq:rhoN}
    \Delta \rho(\omega) = \frac{1}{2\pi i} \partial_\omega \log \det \mathcal{S}^\text{N}(\omega) \qquad , \qquad \det \mathcal{S}^\text{N}(\omega)= \prod_{l=0}^\infty \left(\mathcal{S}^{(1)}_l (\omega)\mathcal{S}^{(2)}_l (\omega)\right)^{D_l^d}
\end{align}
has poles only at the QNM frequencies for the scalar.

\paragraph{Nariai character}

The QNM spectrum for the scalar in question is \cite{Cardoso:2003sw,Molina:2003ff} 
\begin{align}\label{Nariai qnm freq}
	iz_{n,l,+} \ell_N=\Delta_l+n\; ,\quad iz_{n,l,-} \ell_N=\bar{\Delta}_l+n\; ,\quad n=0,1,2,\cdots 
\end{align}
for each $l\geq 0$. These are the poles of \eqref{eq:rhoN}. The QNM character takes the form of a sum of $SO(1,2)$ characters over the KK tower
\begin{align}
	\chi_N (t) \equiv \sum_{n,l,\pm} D_l^d\, e^{-i z_{n,l,\pm}t}=\sum_{l=0}^\infty D_l^d\, \frac{q^{\Delta_l}+q^{\bar{\Delta}_l}}{1-q}
\end{align}
where $q\equiv e^{-t/\ell_N}$. Therefore,
\begin{align}\label{eq:NariaiZbulk}
	\log \widetilde{Z}^\text{Nariai}_\text{bulk}
	=&\int_0^\infty \frac{dt}{2t}\frac{1+q}{1-q}\sum_{l=0}^\infty D_l^d\; \frac{q^{\Delta_l}+q^{\bar{\Delta}_l}}{1-q}\; .
\end{align}

\paragraph{1-loop partition function}

In Euclidean signature, \eqref{eq:narcoords} is Wick-rotated to $S^2\times S^{d-1}$ where $S^2$ and $S^{d-1}$ have radii $\ell_N$ and $r_N$ respectively. We would like to compute the 1-loop determinant for a scalar with mass $m^2$ living on $S^2\times S^{d-1}$ 
\begin{align}\label{eq:nariaiPI}
	Z^\text{N}_\text{PI}=\det \left(-\nabla_N^2+m^2 \right)^{-1/2}\; .
\end{align}
Here the Laplacian $-\nabla^2$ is simply a sum
\begin{align}
	-\nabla_N^2=-\frac{1}{\ell_N^2}\nabla_{S^2}^2-\frac{1}{r_N^2}\nabla_{S^{d-1}}^2\; ,
\end{align}
with eigenvalues and degeneracies
\begin{align}\label{eq:Neigenden}
	\lambda^\text{N}_{p,l}=\lambda^{S^2}_{p}+\lambda^{S^{d-1}}_{l}=\frac{p(p+1)}{\ell_N^2}+\frac{l(l+d-2)}{r_N^2}\quad ,\quad D^\text{N}_{p,l}=D_p^3 \,D_l^d\; .
\end{align}
The path integral \eqref{eq:nariaiPI} has been computed in \cite{Volkov:2000ih} to obtain the semiclassical rate of nucleation of black holes in de Sitter spacetime. 

\begin{figure}[H]
    \centering
   \begin{subfigure}{0.3\textwidth}
            \centering
            \includegraphics[width=\textwidth]{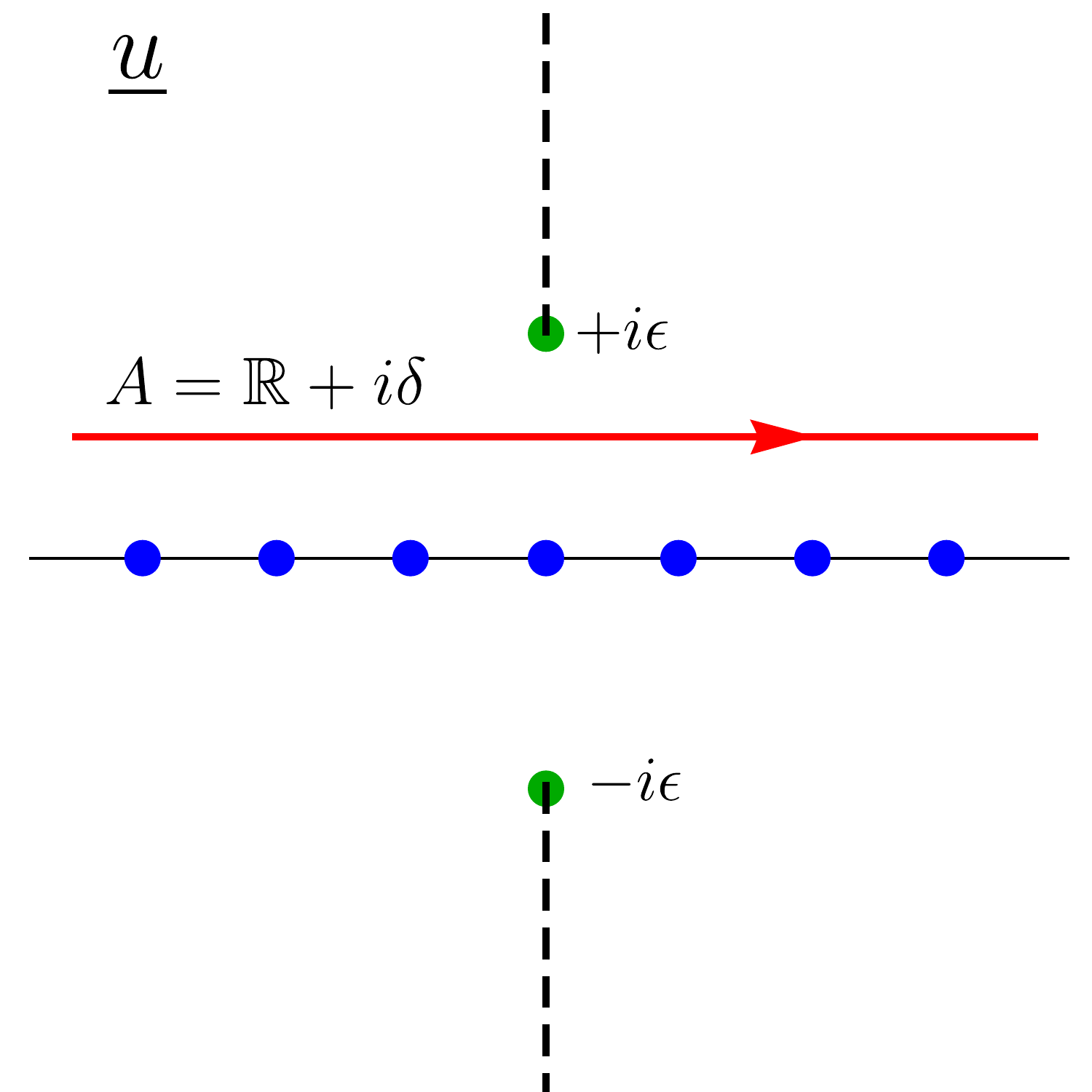}
            \caption[]%
            {{\small original contour}
             }    
        \end{subfigure}
        \begin{subfigure}{0.3\textwidth}  
            \centering 
            \includegraphics[width=\textwidth]{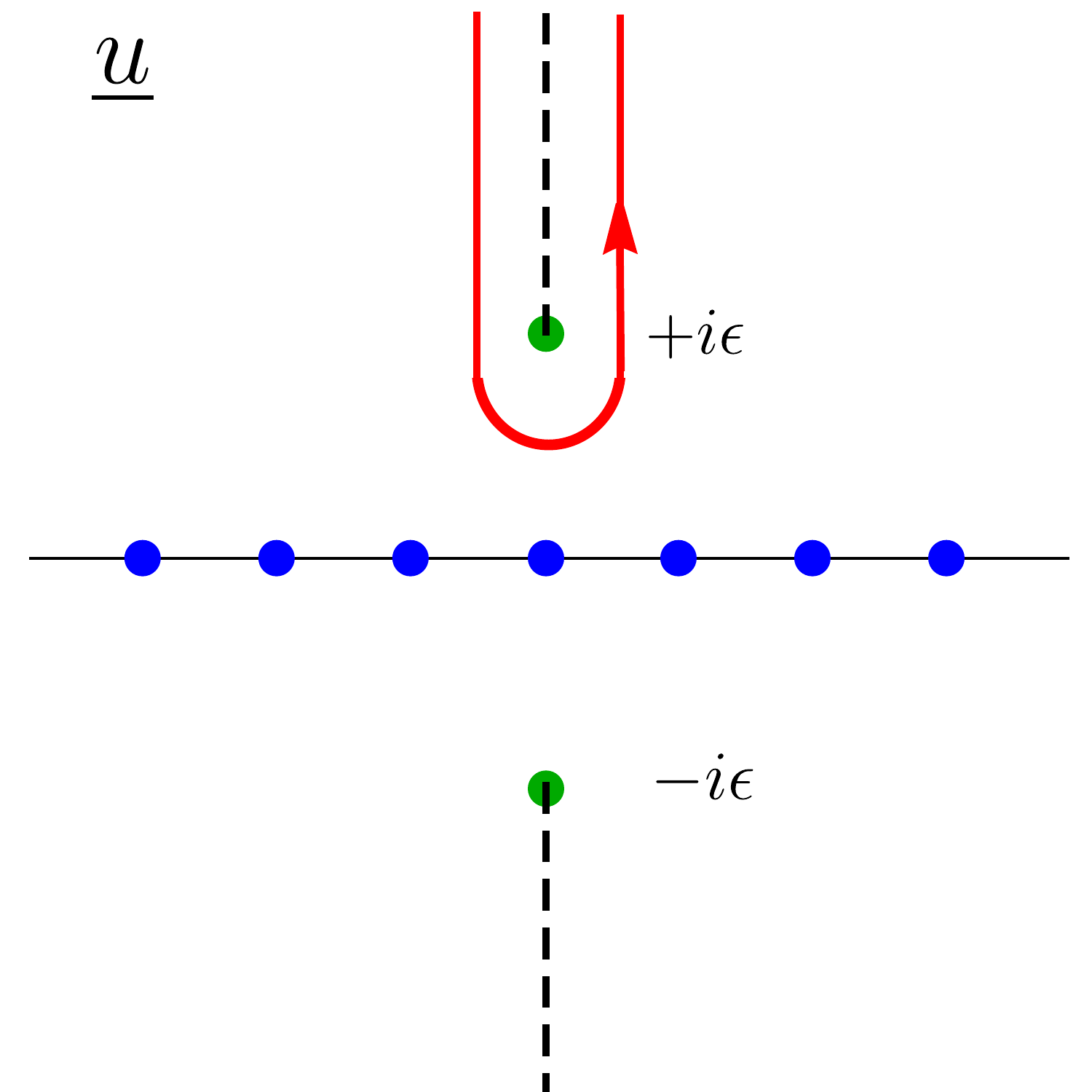}
            \caption[]%
            {{\small \centering  folded contour}}   
        \end{subfigure}
        \begin{subfigure}{0.3\textwidth}  
            \centering 
            \includegraphics[angle=270,width=\textwidth]{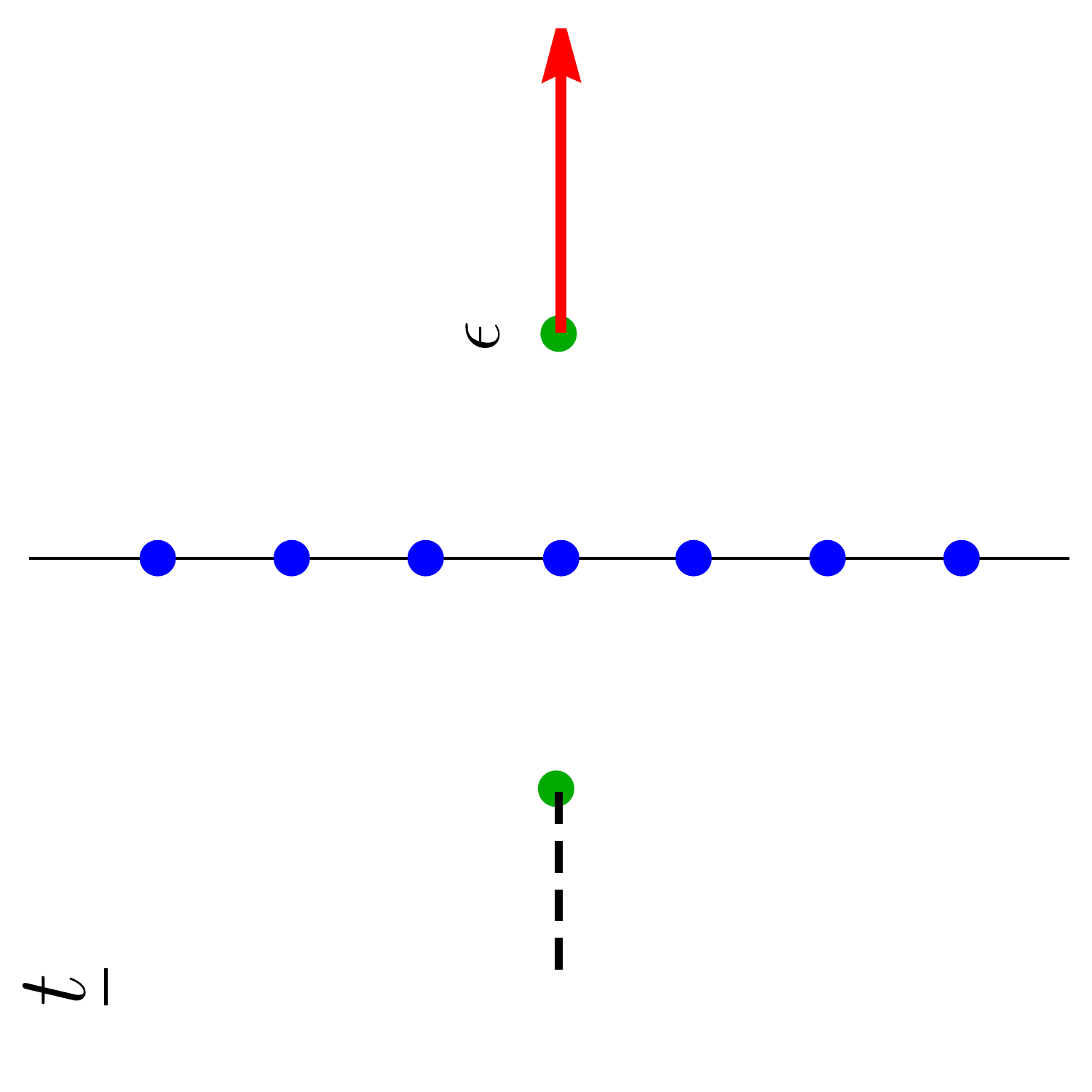}
            \caption[]%
            {{\small \centering  rotated folded contour}}   
        \end{subfigure}
    \caption{We fold the contour $A$ (red) along the branch cut around the branch point $+i \epsilon$ (green dot), and then rotate $u=it$. The blue dots represent the poles of $f(u)$. }\label{fig:hubbard}
\end{figure}

While our discussion in Section \ref{sec:canfn} guarantees that $\log Z^\text{N}_\text{PI}$ must formally agree with the renormalized partition function \eqref{eq:NariaiZbulk}, we would like to demonstrate how to make the UV-regularization more rigorous. To that end, we write \eqref{eq:nariaiPI} in the heat kernel form \cite{Vassilevich:2003xt}
\begin{align}\label{eq:Zpinariai}
	\log Z^\text{N}_\text{PI}=\int_0^\infty \frac{d\tau}{2\tau}e^{-\epsilon^2/4\tau}\sum_{p,l=0}^\infty D_{p,l} \,e^{-\left(\lambda_{p,l}+m^2\right)\tau}\; .
\end{align}
Here we have inserted a regulator $e^{-\epsilon^2/4\tau}$. To proceed, we substitute \eqref{eq:Neigenden} and use the Hubbard-Stratonovich trick (following the approach in \cite{Anninos:2020hfj}) for the sum over $p$ to write
\begin{equation}
	\sum_{p=0}^\infty D_{p}^{3}\, e^{-\tau\left(p+\frac12\right)^2\ell^{-2}_N} = \int_A du \ \frac{e^{-u^2/4\tau}}{\sqrt{4\pi \tau}} f(u) \; ,
\end{equation}
with the integration contour $A = \mathbb{R} + i\delta$, $\delta >0$ (see Fig. \ref{fig:hubbard}). Here we have defined
\begin{equation}
	f(u) \equiv \sum_{p=0}^\infty D_{p}^{3}\, e^{iu\left(p+\frac{1}{2}\right)/\ell_N} = \left(\frac{1+e^{iu/\ell_N}}{1-e^{iu/\ell_N}} \right) \frac{e^{i\frac{u}{2}/\ell_N}}{1-e^{iu/\ell_N}} \; .
\end{equation}
We can then perform the $\tau$-integral in \eqref{eq:Zpinariai} (keeping $\Im u = \delta <\epsilon$). Finally, after deforming the contour $A$ as in Fig. \ref{fig:hubbard} and changing variables to $u = it$, we arrive at the regularized formula
\begin{equation}\label{eq:Nariaireg}
	\log Z_{\text{PI},\epsilon}^\text{N}=\int_\epsilon^\infty \frac{dt}{2\sqrt{t^2-\epsilon^2}}\frac{1+q}{1-q}\sum_{l=0}^\infty D_l^d \frac{e^{-\frac{t}{2\ell_N}+\frac{i}{\ell_N}\nu_l\sqrt{t^2-\epsilon^2}}+e^{-\frac{t}{2\ell_N}-\frac{i}{\ell_N}\nu_l\sqrt{t^2-\epsilon^2}}}{1-q} \; .
\end{equation}
As anticipated, this agrees exactly with \eqref{eq:NariaiZbulk} upon putting $\epsilon=0$. The evaluation of regularized integrals of this form is discussed in \cite{Anninos:2020hfj}.

\section{Black holes in asymptotically flat space and the greybody factor}\label{sec:flat}

In contrast to asymptotically AdS black holes, Hawking radiation emitted by black holes in asymptotically flat space can escape to spatial infinity. Unless the black hole is enclosed by an isolated box, it is not in equilibrium with the radiation gas around it and eventually evaporates. 

Nonetheless, let us comment on some features of the associated scattering problem for free scalars on a fixed black hole background. In this case, we have $F(r) \to 1$ as $r\to\infty$ in \eqref{sym metric}. For example, a $(d+1)$-dimensional Schwarzschild has $F(r) = 1- \frac{C}{r^{d-2}}$. The Klein-Gordon equation can still be rewritten in the form \eqref{eq:eff Sch}, except that there will be two asymptotic regions:
\begin{align}\label{flat regions}
	\text{Spatial infinity: } \quad x \to \infty \qquad \text{Horizon: } \quad x \to -\infty\;.
\end{align}
In other words, this is a two-channel scattering problem, in sharp contrast with the AdS case. Moreover, a mass term $m^2$ would lead to a finite gap: $V_l(+\infty)-V_l(-\infty)=m^2$; waves sent from the horizon with energy $\omega^2<m^2$ are totally reflected back to the black hole. To avoid such complications we focus on the massless case. A solution to \eqref{eq:eff Sch} has the asymptotic behavior
\begin{align}
    \psi_l(x) =
    \begin{cases}
        A_l^\text{in} (\omega)\, e^{i\omega x} + A_l^\text{out} (\omega)\, e^{-i\omega x} &, \quad  x\to -\infty \\
        B_l^\text{out} (\omega) \, e^{i\omega x}+ B_l^\text{in}  (\omega)\, e^{-i \omega x}&, \quad x\to \infty
    \end{cases} \; .
\end{align}
The S-matrix maps the incoming coefficients to the outgoing coefficients, i.e.
\begin{align}\label{eq:flat S}
	\begin{pmatrix}
		A_l^\text{out}(\omega) \\
		B_l^\text{out}(\omega)
	\end{pmatrix}
	=
    \mathcal{S}_l(\omega)
	\begin{pmatrix}
		A_l^\text{in} (\omega)\\
		B_l^\text{in}(\omega)
	\end{pmatrix} \; , \qquad \mathcal{S}_l(\omega)=
	\begin{pmatrix}
		R_l(\omega) & T_l(\omega)\\
		T'_l(\omega) & R'_l(\omega) 
	\end{pmatrix} \; .
\end{align}
Here we have introduced the reflection ($R,R'$) and transmission ($T,T'$) coefficients. These coefficients are not independent. First, since \eqref{eq:eff Sch} is invariant upon $\omega \to - \omega$, we have $A_l^\text{in} (\omega)=A_l^\text{out} (-\omega)$ and $B_l^\text{in} (\omega)=B_l^\text{out} (-\omega)$. Sending $\omega \to -\omega$ in \eqref{eq:flat S}, we deduce
\begin{align}\label{eq:Sminusw}
    \mathcal{S}_l(-\omega)\mathcal{S}_l(\omega)=\mathcal{S}_l(\omega)\mathcal{S}_l(-\omega) = I \; .
\end{align}
Second, using \eqref{eq:eff Sch} one can check that the current 
\begin{align}\label{eq:conJ}
    J_{\omega l} (x) = \psi_{-\omega l} \partial_x \psi_{\omega l}-\psi_{\omega l} \partial_x \psi_{-\omega l}
\end{align}
is conserved: $\partial_x J_{\omega l} (x) =0$, which leads to another constraint
\begin{align}\label{eq:Sunitary}
    \mathcal{S}_l^T(-\omega)\mathcal{S}_l(\omega)=\mathcal{S}_l(\omega)\mathcal{S}_l^T(-\omega) = I \; .
\end{align}
Conditions \eqref{eq:Sminusw} and \eqref{eq:Sunitary} together imply
\begin{align}\label{R T relation}
	T'_l(\omega)= T_l(\omega) \qquad \text{and} \qquad R_l(-\omega) T_l(\omega)+R'_l(\omega)T_l(-\omega)= 0 \;.
\end{align}
When $\omega$ is real, we have $\left(A_l^\text{in} (\omega)\right)^*=A_l^\text{out} (\omega)$ and $\left(B_l^\text{in} (\omega)\right)^*=B_l^\text{out} (\omega)$; the conserved current \eqref{eq:conJ} is same as the probability current; the condition \eqref{eq:Sunitary} is same as saying $\mathcal{S}$ is unitary. 

\paragraph{The phase and magnitude of the transmission coefficient}

Similar to the single-channel scattering discussed in Section \ref{sec:char DOS}, the DOS $\rho_l(\omega)$ is infinite but its change $\Delta \rho_l(\omega)$ relative to some reference problem is finite. The relation \eqref{friedel relation cos} is generalized to \cite{PhysRevB.32.2674}
\begin{align}\label{friedel relation}
	\Delta \rho_l(\omega)=\rho_l(\omega)-\bar\rho_l(\omega) =  \frac{1}{2\pi i} \partial_\omega \tr \left(\log \mathcal{S}_l(\omega) - \log \bar{\mathcal{S}}_l(\omega)\right) \; .
\end{align}
Here tr is the trace over the 2 by 2 matrix \eqref{eq:flat S}. For a flat reference potential $\bar V(x)=0$, $\bar{\mathcal{S}}_l$ is simply the identity matrix. Using \eqref{R T relation}, one can show that for real $\omega$, $\tr \log \mathcal{S}_l$ is essentially the phase $\theta_T$ of the transmission coefficient $T=|T|e^{i\theta_T}$. This establishes a pleasing connection with another quantity of interest in the study of black hole thermodynamics: the greybody factor \cite{Hawking:1975vcx}
\begin{align}
	\gamma_l^\text{greybody} (\omega) \equiv T_l(\omega) T_l(-\omega) = |T_l(\omega)|^2 \; .
\end{align}
Therefore, for real $\omega$, the magnitude of $T_l$ measures the absorption/transmission probability, while its phase captures information about the DOS.


\section{Discussion and outlook}\label{sec:discussion}

To conclude, we have provided evidence for the Lorentzian interpretation of the Euclidean path integral through the manifestly covariant relation \eqref{eq:ratioPI}. With the switch of perspective to that of 1D scattering, it is natural to expect more insights could be imported from scattering theory in open quantum systems (see for instance \cite{ChaosBook}) into understanding the quantum structures of black holes. On the Lorentzian side, the discussion in Section \ref{sec:char DOS} may indicate a (non-perturbative) scattering formulation generalizing \eqref{eq:ratioPI} to interacting QFTs on a fixed black hole background, perhaps along the lines of \cite{PhysRev.187.345}. Establishing the equality \eqref{eq:ratioPI} rigorously will likely involve carefully cutting and gluing the Euclidean path integral around the origin.


From the point of view of the global two-sided geometry (see Figure \ref{fig:penrose scatter}), the starting point \eqref{log Z bulk} of our Lorentzian calculation can be viewed as computing the normalization of the reduced density matrix obtained by tracing out the Hartle-Hawking state along half of the spatial slice. This assumes that the global Hilbert space factorizes. From an algebraic viewpoint (reviewed in \cite{Witten:2018zxz}), such a factorization does not actually exist; the algebra of observables for the scalar QFT in the outside-horizon region is a Type III von Neumann algebra, which does not admit a trace. The infinity of the single-particle DOS $\rho(\omega)$ in \eqref{log Z bulk}  can be viewed as a manifestation of this non-factorization of Hilbert space. As explained in \cite{Witten:2021jzq}, Type III algebras also arise when describing the thermodynamic or large volume limit of a system directly in terms of operators acting on a Hilbert space; indeed, with the scattering picture in Section \ref{sec:scattering}, we are viewing the free scalar QFT as an infinite-volume quantum statistical system, and $\rho(\omega)$ diverges precisely due to the infinite size of the box. 

As pointed out recently in \cite{Witten:2021unn,Chandrasekaran:2022cip}, including 1-loop corrections from gravity, the algebra for the scalar QFT outside the horizon turns from Type III to Type II, for which a trace can be defined up to an arbitrary (infinite) normalization. In our field theory calculation, extracting non-trivial information from $\log Z_\text{bulk}$ involves an arbitrary choice of reference $\log \bar Z_\text{bulk}$ as well. Even though we have not discussed in detail, gravity is indeed crucial for eventually making physical quantities UV-finite in the framework of low-energy effective theory of gravity plus matter. Integrals such as \eqref{eq:DHSintegral} are typically UV-divergent and require regularization. An example regulated by the heat kernel method is given in \eqref{eq:Nariaireg}; such an integral will have the structure $\log Z_{\text{PI},\epsilon}= \log Z^\text{UV}_{\text{PI},\epsilon}+\log Z^\text{finite}_{\text{PI}}$, where the UV-divergent part takes the form
\begin{align}\label{eq:uvdiv}
    \log Z^\text{UV}_{\text{PI},\epsilon}=\sum_{k=0}^{d}\frac{B_k}{\epsilon^{d+1-k}}+B_{d+1}\log \frac{L}{\epsilon} \; ,
\end{align}
where $L$ is a parameter with a dimension of length. Here $B_k$ are related to heat kernel coefficients \cite{Vassilevich:2003xt}, which can be expressed in terms of curvature invariants on the manifold.\footnote{For manifolds without a boundary, $B_k=0$ for odd $k$.} Coupling the theory to gravity, all the UV-divergences \eqref{eq:uvdiv} will be absorbed into the renormalization of the cosmological constant, Newton's constant and higher-curvature couplings in the effective gravitational action, after which we are left with a UV-finite quantity $\log Z^\text{finite}_{\text{PI}}$. In light of these suggestive observations, it would be extremely interesting to investigate the precise connection between the direct field theory and formal algebraic approaches.

The central role of QNMs in the DHS formula raises another interesting prospect of studying the thermal or entanglement properties of astrophysical black holes by probing their QNMs, for instance through gravitational-wave ringdown \cite{Berti:2009kk} or photon ring \cite{Hadar:2022xag} measurements. QNM frequencies are in general difficult to compute exactly; however, their asymptotic forms in certain regimes, for instance high-overtone ($n\to \infty$) or eikonal ($l\to \infty$), are often analytically computable (reviewed in \cite{Berti:2009kk,Konoplya:2011qq}). One may be able to extract useful information about black holes by combining these approximations with our formula \eqref{eq:DHSintegral}. 


As far as microscopic models are concerned, in AdS/CFT it is known that QNMs describe the decay of perturbations in the dual CFT and appear as poles of the boundary retarded Green’s functions \cite{Son:2002sd,Horowitz:1999jd,Birmingham:2001pj}; in the context of string theory, given the success of the Euclidean gravity method in reproducing the microscopic counting of black hole entropies \cite{ Banerjee:2010qc, Banerjee:2011jp, Sen:2012kpz, Sen:2014aja}, it seems possible through the DHS formula \eqref{eq:DHSintegral} to identify a description of QNMs in terms of microscopic degrees of freedom. Formulas of the form \eqref{eq:DHSintegral} have proved useful for theories with an infinite tower of fields. Instead of calculating the 1-loop determinant one by one before summing over the spectrum, one could sum the QNM characters first before computing the integral. For example, as demonstrated in \cite{Anninos:2020hfj,Sun:2020ame}, 1-loop tests for Higher Spin AdS/CFT \cite{Giombi:2013fka,Giombi:2014iua,Giombi:2016pvg,Gunaydin:2016amv} can be performed in a much more compact manner. 

Finally, for some of the above-mentioned and other applications one needs to generalize our considerations to more general black hole backgrounds (ones with spins or other charges, for example incorporating a background electric field along the lines of \cite{Grewal:2021bsu}). We leave this to future work.


\section*{Acknowledgments} 

It is a great pleasure to thank Daniel Jafferis, Subir Sachdev, Andy Strominger and Gabriel Wong for stimulating conversations, and especially Dionysios Anninos, Frederik Denef, Manvir Grewal, Temple He, Daniel Kapec and Zimo Sun for useful discussions and comments on the draft. AL  was supported in part by the Croucher Foundation and U.S. Department of Energy grant de-sc0007870. KP was supported in part by the U.S. Department of Energy grant de-sc0011941.

\appendix

\section{Scattering in the Rindler-like region}\label{app:rindler}

In this appendix we study the scattering problem \eqref{eq:nearhorscatt} (with all the primes dropped)
\begin{align}\label{appeq:rindlerscatt}
    \left[ -\partial_{x}^2 + \left(\frac{4\pi}{\beta} \right)^2 e^{\frac{4\pi}{\beta}x} \right] \psi(x) = \omega^2 \psi(x) \; .
\end{align}
As explained in Section \ref{sec:canfn}, this is relevant to an observer probing the region near a black hole horizon of temperature $\beta$. The general solution to \eqref{appeq:rindlerscatt} is a linear combination of modified Bessel functions
\begin{align}
    \psi (x) = C_\text{n.} K_{\frac{i \beta  \omega }{2 \pi }}\left(2 e^{\frac{2 \pi  x}{\beta }}\right) +C_\text{n.n.} I_{\frac{i \beta  \omega }{2 \pi }}\left(2 e^{\frac{2 \pi  x}{\beta }}\right) \; .
\end{align}
Here n. and n.n. means the solutions are respectively normalizable and non-normalizable, in the sense that they are exponentially decaying/growing as $x \to \infty$.\footnote{As $z\to \infty$, $K_{\alpha}\left( z\right)\propto \frac{1}{\sqrt{z}}e^{-z}$ and $I_{\alpha}\left( z\right)\propto \frac{1}{\sqrt{z}}e^{z}$.} Intuitively, the problem \eqref{appeq:rindlerscatt} is similar to that in the asymptotically AdS case: there is a infinite potential well at spatial infinity, except that the potential well here is due to the acceleration of the observer. We impose a Dirichlet boundary condition: $C_\text{n.n.}=0$. Near the Rindler horizon, the normalizable solution behaves as
\begin{align}\label{eq:Rindler near hor}
   \psi (x\to -\infty) \propto  \Gamma
   \left(\frac{i \beta  \omega }{2 \pi }\right) e^{-i \omega x}+\Gamma
   \left(-\frac{i \beta  \omega }{2 \pi }\right) e^{i \omega x} \; .
\end{align}
The ratio between the coefficients of the outgoing and incoming waves defines a unitary S-matrix:
\begin{align}\label{appeq:RindlerS}
    \mathcal{S}^\text{Rindler} (\beta,\omega) = \frac{\Gamma
   \left(\frac{i \beta  \omega }{2 \pi }\right)}{\Gamma
   \left(-\frac{i \beta  \omega }{2 \pi }\right)} \; .
\end{align}
Notice that the S-matrix hits a pole or zero whenever $\omega$ meets the Matsubara frequencies
\begin{align}\label{eq:RindlerQNM}
    \omega = \omega_{\pm , n} = \pm i \frac{2\pi n}{\beta} \; , \qquad n = 1 , 2 , 3 ,\cdots  \; ,
\end{align}
at which the mode function behaves like
\begin{align}
    K_{\mp n}\left(2e^{\frac{2 \pi  x}{\beta }}\right) \propto e^{- \frac{2\pi n}{\beta} x} \;, \qquad x\to -\infty \; .
\end{align}
Therefore, the $\pm$ (quasinormal) modes \eqref{eq:RindlerQNM} are purely incoming (outgoing).  


\section{Scalar on global \texorpdfstring{$AdS_3$}{}}

Even though global $AdS_3$ (setting $\ell_\text{AdS}=1$)
\begin{align}\label{appeq:ads3}
    ds^2 = -\left(1+r^2 \right) dt^2 + \frac{dr^2}{1+r^2} +r^2 d\phi^2 
\end{align}
does not have a horizon and the considerations in this paper do not apply, this example is closely related to the BTZ case. Moreover, it is instructive to highlight the difference between the two computations.


The {\it normal} mode spectrum for a scalar with mass $m^2= \Delta (\Delta-2)$ on \eqref{appeq:ads3} is well-known:
\begin{align}\label{appeq:adsnormal}
	\omega_{n,p}= 2n+|l|+\Delta
\end{align}
where $n=0,1,2,\dots$ is the ``overtone" number and $l=0,\pm 1,\pm 2, \dots$ labels the $U(1)$ angular momentum quantum number. The DOS is simply a sum of delta functions over the discrete spectrum \eqref{appeq:adsnormal}. The thermal canonical partition function is
\begin{align}
	\log Z_\text{bulk}^{AdS_3}\equiv \log \Tr \, e^{-\beta \hat{H}} =-\sum_{n,l} \left(\log (1-e^{-\beta\omega_{n,l}}) +\frac{\beta \omega_{n,l}}{2}\right) \; .
\end{align}
The second term in the bracket is an infinite contribution from zero point energies, which renormalizes the cosmological constant. We will drop this from now on. Expanding the logarithm as a series and performing the sums over $n$ and $l$, we have
\begin{align}\label{appeq:TAdS}
	\log Z_\text{bulk}^{AdS_3}=\sum_{k=1}^\infty \frac{1}{k}\frac{e^{-\Delta k \beta}}{(1-e^{-k \beta})^2}=\sum_{k=1}^\infty \frac{\chi^{AdS_{3}}(k \beta)}{k}\; , \qquad \chi^{AdS_{3}}(t)=\frac{e^{-\Delta t}}{(1-e^{-t})^2}\; .
\end{align}
In the last equality we have expressed the result in terms of the $SO(2,2)$ character $\chi^{AdS_{3}}(t)$ (see for example \cite{Dolan:2005wy}). This result has been computed in \cite{Giombi:2008vd} using the Euclidean path integral on $TAdS_3$.

\section{Example: Scalar on de Sitter static patch}\label{app:dS}

We consider a scalar with mass $m^2 \ell_\text{dS}^2=\Delta \bar\Delta\equiv \Delta (d-\Delta)$ living on a static patch in $dS_{d+1}$
\begin{align}\label{appeq:dSmetric}
    ds^2 = -\left(1-\frac{r^2}{\ell_\text{dS}^2} \right) dt^2 + \frac{dr^2}{1-\frac{r^2}{\ell_\text{dS}^2}} +r^2 d\phi^2 \quad ,\qquad  0\leq r < \ell_\text{dS} \quad , \qquad \ell_\text{dS} \equiv \sqrt{\frac{d(d-1)}{2\Lambda}}  \; .
\end{align}
The de Sitter horizon is at $r=\ell_\text{dS}$, with temperature $T_\text{dS}=\frac{1}{2\pi \ell_\text{dS}}$. This is the context where the relevance of the scattering picture in understanding the Euclidean path integral was first pointed out (see Appendix B.3 of \cite{Anninos:2020hfj}), which directly inspired the current work. In terms of the tortoise coordinate $x=\int_0^r \frac{dr'}{1-{r'}^2/\ell_\text{dS}^2}=\ell_\text{dS} \tanh^{-1}\frac{r}{\ell_\text{dS}}$, solving $(-\nabla^2+m^2) \phi =0$ on \eqref{appeq:dSmetric} with the ansatz \eqref{eq:normalmodes} while imposing the regularity condition at the location of the observer ($r=0$), one finds the near-horizon behavior 
\begin{align}
    \psi_l(x\to \infty) \propto & \frac{\Gamma\left(i \ell_\text{dS} \omega\right)}{\Gamma\left(\frac{\Delta+l+i \ell_\text{dS} \omega}{2}\right) \Gamma\left(\frac{\bar\Delta+l+ i \ell_\text{dS} \omega}{2}\right)} e^{i \omega x}+\frac{\Gamma\left(-i \ell_\text{dS} \omega\right)}{\Gamma\left(\frac{\Delta+l- i \ell_\text{dS} \omega}{2}\right) \Gamma\left(\frac{\bar\Delta+l- i \ell_\text{dS} \omega}{2}\right)} e^{-i \omega x} \;.
\end{align}
Therefore, the S-matrix $\mathcal{S}_l (\omega)$ has the same general structure as \eqref{eq:btzsmatrix} and \eqref{eq:nariaismatrix}, that is, $\mathcal{S}_l (\omega)  = \mathcal{S}_l^\text{dS} (\omega) \mathcal{S}^\text{Rindler} \left(2\pi \ell_\text{dS},\omega\right)$, where $\mathcal{S}^\text{Rindler} \left(\beta,\omega\right)$ is the Rindler S-matrix \eqref{eq:RindlerS} and
\begin{gather}
   \mathcal{S}_l^\text{dS} (\omega)\equiv \frac{\Gamma\left(\frac{\Delta+l- i \ell_\text{dS} \omega}{2}\right) \Gamma\left(\frac{\bar\Delta+l- i \ell_\text{dS} \omega}{2}\right)}{\Gamma\left(\frac{\Delta+l+i \ell_\text{dS} \omega}{2}\right) \Gamma\left(\frac{\bar\Delta+l+ i \ell_\text{dS} \omega}{2}\right)}  \; 
\end{gather}
captures all the QNMs. We refer the readers to \cite{Anninos:2020hfj} for an elaborate discussion (including a careful treatment of UV-regularization) on the Lorentzian and Euclidean (sphere) partition functions in this context.


\bibliographystyle{utphys}
\bibliography{ref}

\providecommand{\href}[2]{#2}\begingroup\raggedright\begin{thebibliography}{10}

\bibitem{Gibbons:1976ue}
G.~W. Gibbons and S.~W. Hawking, ``{Action Integrals and Partition Functions in
  Quantum Gravity},'' \href{http://dx.doi.org/10.1103/PhysRevD.15.2752}{{\em
  Phys. Rev. D} {\bfseries 15} (1977) 2752--2756}.

\bibitem{Banerjee:2010qc}
S.~Banerjee, R.~K. Gupta, and A.~Sen, ``{Logarithmic Corrections to Extremal
  Black Hole Entropy from Quantum Entropy Function},''
  \href{http://dx.doi.org/10.1007/JHEP03(2011)147}{{\em JHEP} {\bfseries 03}
  (2011) 147}, \href{http://arxiv.org/abs/1005.3044}{{\ttfamily arXiv:1005.3044
  [hep-th]}}.

\bibitem{Banerjee:2011jp}
S.~Banerjee, R.~K. Gupta, I.~Mandal, and A.~Sen, ``{Logarithmic Corrections to
  N=4 and N=8 Black Hole Entropy: A One Loop Test of Quantum Gravity},''
  \href{http://dx.doi.org/10.1007/JHEP11(2011)143}{{\em JHEP} {\bfseries 11}
  (2011) 143}, \href{http://arxiv.org/abs/1106.0080}{{\ttfamily arXiv:1106.0080
  [hep-th]}}.

\bibitem{Sen:2012dw}
A.~Sen, ``{Logarithmic Corrections to Schwarzschild and Other Non-extremal
  Black Hole Entropy in Different Dimensions},''
  \href{http://dx.doi.org/10.1007/JHEP04(2013)156}{{\em JHEP} {\bfseries 04}
  (2013) 156}, \href{http://arxiv.org/abs/1205.0971}{{\ttfamily arXiv:1205.0971
  [hep-th]}}.

\bibitem{Sen:2012kpz}
A.~Sen, ``{Logarithmic Corrections to N=2 Black Hole Entropy: An Infrared
  Window into the Microstates},''
  \href{http://dx.doi.org/10.1007/s10714-012-1336-5}{{\em Gen. Rel. Grav.}
  {\bfseries 44} no.~5, (2012) 1207--1266},
  \href{http://arxiv.org/abs/1108.3842}{{\ttfamily arXiv:1108.3842 [hep-th]}}.

\bibitem{Sen:2014aja}
A.~Sen, ``{Microscopic and Macroscopic Entropy of Extremal Black Holes in
  String Theory},'' \href{http://dx.doi.org/10.1007/s10714-014-1711-5}{{\em
  Gen. Rel. Grav.} {\bfseries 46} (2014) 1711},
  \href{http://arxiv.org/abs/1402.0109}{{\ttfamily arXiv:1402.0109 [hep-th]}}.

\bibitem{tHooft:1984kcu}
G.~'t~Hooft, ``{On the Quantum Structure of a Black Hole},''
  \href{http://dx.doi.org/10.1016/0550-3213(85)90418-3}{{\em Nucl. Phys. B}
  {\bfseries 256} (1985) 727--745}.

\bibitem{Frolov:1998vs}
V.~P. Frolov and D.~V. Fursaev, ``{Thermal fields, entropy, and black holes},''
  \href{http://dx.doi.org/10.1088/0264-9381/15/8/001}{{\em Class. Quant. Grav.}
  {\bfseries 15} (1998) 2041--2074},
  \href{http://arxiv.org/abs/hep-th/9802010}{{\ttfamily arXiv:hep-th/9802010}}.

\bibitem{Solodukhin:2011gn}
S.~N. Solodukhin, ``{Entanglement entropy of black holes},''
  \href{http://dx.doi.org/10.12942/lrr-2011-8}{{\em Living Rev. Rel.}
  {\bfseries 14} (2011) 8}, \href{http://arxiv.org/abs/1104.3712}{{\ttfamily
  arXiv:1104.3712 [hep-th]}}.

\bibitem{Anninos:2020hfj}
D.~Anninos, F.~Denef, Y.~T.~A. Law, and Z.~Sun, ``{Quantum de Sitter horizon
  entropy from quasicanonical bulk, edge, sphere and topological string
  partition functions},'' \href{http://dx.doi.org/10.1007/JHEP01(2022)088}{{\em
  JHEP} {\bfseries 01} (2022) 088},
  \href{http://arxiv.org/abs/2009.12464}{{\ttfamily arXiv:2009.12464
  [hep-th]}}.

\bibitem{ChaosBook}
P.~Cvitanovic, R.~Artuso, R.~Mainieri, G.~Tanner, and G.~Vattay, {\em Chaos:
  Classical and Quantum}.
\newblock Niels Bohr Inst., Copenhagen, 2016.
\newblock \url{http://ChaosBook.org/}.

\bibitem{Denef:2009kn}
F.~Denef, S.~A. Hartnoll, and S.~Sachdev, ``{Black hole determinants and
  quasinormal modes},''
  \href{http://dx.doi.org/10.1088/0264-9381/27/12/125001}{{\em Class. Quant.
  Grav.} {\bfseries 27} (2010) 125001},
  \href{http://arxiv.org/abs/0908.2657}{{\ttfamily arXiv:0908.2657 [hep-th]}}.

\bibitem{Susskind:1994sm}
L.~Susskind and J.~Uglum, ``{Black hole entropy in canonical quantum gravity
  and superstring theory},''
  \href{http://dx.doi.org/10.1103/PhysRevD.50.2700}{{\em Phys. Rev. D}
  {\bfseries 50} (1994) 2700--2711},
  \href{http://arxiv.org/abs/hep-th/9401070}{{\ttfamily arXiv:hep-th/9401070}}.

\bibitem{Demers:1995dq}
J.-G. Demers, R.~Lafrance, and R.~C. Myers, ``{Black hole entropy without brick
  walls},'' \href{http://dx.doi.org/10.1103/PhysRevD.52.2245}{{\em Phys. Rev.
  D} {\bfseries 52} (1995) 2245--2253},
  \href{http://arxiv.org/abs/gr-qc/9503003}{{\ttfamily arXiv:gr-qc/9503003}}.

\bibitem{Witten:2018zxz}
E.~Witten, ``{APS Medal for Exceptional Achievement in Research: Invited
  article on entanglement properties of quantum field theory},''
  \href{http://dx.doi.org/10.1103/RevModPhys.90.045003}{{\em Rev. Mod. Phys.}
  {\bfseries 90} no.~4, (2018) 045003},
  \href{http://arxiv.org/abs/1803.04993}{{\ttfamily arXiv:1803.04993
  [hep-th]}}.

\bibitem{Witten:2021unn}
E.~Witten, ``{Gravity and the Crossed Product},''
  \href{http://arxiv.org/abs/2112.12828}{{\ttfamily arXiv:2112.12828
  [hep-th]}}.

\bibitem{Chandrasekaran:2022cip}
V.~Chandrasekaran, R.~Longo, G.~Penington, and E.~Witten, ``{An Algebra of
  Observables for de Sitter Space},''
  \href{http://arxiv.org/abs/2206.10780}{{\ttfamily arXiv:2206.10780
  [hep-th]}}.

\bibitem{BHedge}
M.~Grewal, Y.~T.~A. Law, and K.~Parmentier, ``{Black hole horizon edge
  partition functions},'' \href{http://arxiv.org/abs/2207.XXXXX}{{\ttfamily
  arXiv:2207.XXXXX [hep-th]}}.

\bibitem{Law:2020cpj}
Y.~T.~A. Law, ``{A compendium of sphere path integrals},''
  \href{http://dx.doi.org/10.1007/JHEP12(2021)213}{{\em JHEP} {\bfseries 12}
  (2021) 213}, \href{http://arxiv.org/abs/2012.06345}{{\ttfamily
  arXiv:2012.06345 [hep-th]}}.

\bibitem{Vassilevich:2003xt}
D.~V. Vassilevich, ``{Heat kernel expansion: User's manual},''
  \href{http://dx.doi.org/10.1016/j.physrep.2003.09.002}{{\em Phys. Rept.}
  {\bfseries 388} (2003) 279--360},
  \href{http://arxiv.org/abs/hep-th/0306138}{{\ttfamily arXiv:hep-th/0306138}}.

\bibitem{Hawking:1976ja}
S.~W. Hawking, ``{Zeta Function Regularization of Path Integrals in Curved
  Space-Time},'' \href{http://dx.doi.org/10.1007/BF01626516}{{\em Commun. Math.
  Phys.} {\bfseries 55} (1977) 133}.

\bibitem{Jafferis:2015del}
D.~L. Jafferis, A.~Lewkowycz, J.~Maldacena, and S.~J. Suh, ``{Relative entropy
  equals bulk relative entropy},''
  \href{http://dx.doi.org/10.1007/JHEP06(2016)004}{{\em JHEP} {\bfseries 06}
  (2016) 004}, \href{http://arxiv.org/abs/1512.06431}{{\ttfamily
  arXiv:1512.06431 [hep-th]}}.

\bibitem{Son:2002sd}
D.~T. Son and A.~O. Starinets, ``{Minkowski space correlators in AdS / CFT
  correspondence: Recipe and applications},''
  \href{http://dx.doi.org/10.1088/1126-6708/2002/09/042}{{\em JHEP} {\bfseries
  09} (2002) 042}, \href{http://arxiv.org/abs/hep-th/0205051}{{\ttfamily
  arXiv:hep-th/0205051}}.

\bibitem{Anninos:2011af}
D.~Anninos, S.~A. Hartnoll, and D.~M. Hofman, ``{Static Patch Solipsism:
  Conformal Symmetry of the de Sitter Worldline},''
  \href{http://dx.doi.org/10.1088/0264-9381/29/7/075002}{{\em Class. Quant.
  Grav.} {\bfseries 29} (2012) 075002},
  \href{http://arxiv.org/abs/1109.4942}{{\ttfamily arXiv:1109.4942 [hep-th]}}.

\bibitem{Ching:1994bd}
E.~S.~C. Ching, P.~T. Leung, W.~M. Suen, and K.~Young, ``{Late time tail of
  wave propagation on curved space-time},''
  \href{http://dx.doi.org/10.1103/PhysRevLett.74.2414}{{\em Phys. Rev. Lett.}
  {\bfseries 74} (1995) 2414--2417},
  \href{http://arxiv.org/abs/gr-qc/9410044}{{\ttfamily arXiv:gr-qc/9410044}}.

\bibitem{Ching:1995tj}
E.~S.~C. Ching, P.~T. Leung, W.~M. Suen, and K.~Young, ``{Wave propagation in
  gravitational systems: Late time behavior},''
  \href{http://dx.doi.org/10.1103/PhysRevD.52.2118}{{\em Phys. Rev. D}
  {\bfseries 52} (1995) 2118--2132},
  \href{http://arxiv.org/abs/gr-qc/9507035}{{\ttfamily arXiv:gr-qc/9507035}}.

\bibitem{PhysRevD.5.2419}
R.~H. Price, ``Nonspherical perturbations of relativistic gravitational
  collapse. i. scalar and gravitational perturbations,''
  \href{http://dx.doi.org/10.1103/PhysRevD.5.2419}{{\em Phys. Rev. D}
  {\bfseries 5} (May, 1972) 2419--2438}.
  \url{https://link.aps.org/doi/10.1103/PhysRevD.5.2419}.

\bibitem{Mann:1996ze}
R.~B. Mann and S.~N. Solodukhin, ``{Quantum scalar field on three-dimensional
  (BTZ) black hole instanton: Heat kernel, effective action and
  thermodynamics},'' \href{http://dx.doi.org/10.1103/PhysRevD.55.3622}{{\em
  Phys. Rev. D} {\bfseries 55} (1997) 3622--3632},
  \href{http://arxiv.org/abs/hep-th/9609085}{{\ttfamily arXiv:hep-th/9609085}}.

\bibitem{Cardoso:2001hn}
V.~Cardoso and J.~P.~S. Lemos, ``{Scalar, electromagnetic and Weyl
  perturbations of BTZ black holes: Quasinormal modes},''
  \href{http://dx.doi.org/10.1103/PhysRevD.63.124015}{{\em Phys. Rev. D}
  {\bfseries 63} (2001) 124015},
  \href{http://arxiv.org/abs/gr-qc/0101052}{{\ttfamily arXiv:gr-qc/0101052}}.

\bibitem{1950SRToh..34..160N}
H.~{Nariai}, ``{On some static solutions of Einstein's gravitational field
  equations in a spherically symmetric case},'' {\em Sci. Rep. Tohoku Univ.
  Eighth Ser.} {\bfseries 34} (Jan., 1950) 160.

\bibitem{Susskind:2021dfc}
L.~Susskind, ``{Black Holes Hint Towards De Sitter-Matrix Theory},''
  \href{http://arxiv.org/abs/2109.01322}{{\ttfamily arXiv:2109.01322
  [hep-th]}}.

\bibitem{Cardoso:2003sw}
V.~Cardoso and J.~P.~S. Lemos, ``{Quasinormal modes of the near extremal
  Schwarzschild-de Sitter black hole},''
  \href{http://dx.doi.org/10.1103/PhysRevD.67.084020}{{\em Phys. Rev. D}
  {\bfseries 67} (2003) 084020},
  \href{http://arxiv.org/abs/gr-qc/0301078}{{\ttfamily arXiv:gr-qc/0301078}}.

\bibitem{Molina:2003ff}
C.~Molina, ``{Quasinormal modes of d-dimensional spherical black holes with
  near extreme cosmological constant},''
  \href{http://dx.doi.org/10.1103/PhysRevD.68.064007}{{\em Phys. Rev. D}
  {\bfseries 68} (2003) 064007},
  \href{http://arxiv.org/abs/gr-qc/0304053}{{\ttfamily arXiv:gr-qc/0304053}}.

\bibitem{Volkov:2000ih}
M.~S. Volkov and A.~Wipf, ``{Black hole pair creation in de Sitter space: A
  Complete one loop analysis},''
  \href{http://dx.doi.org/10.1016/S0550-3213(00)00287-X}{{\em Nucl. Phys. B}
  {\bfseries 582} (2000) 313--362},
  \href{http://arxiv.org/abs/hep-th/0003081}{{\ttfamily arXiv:hep-th/0003081}}.

\bibitem{PhysRevB.32.2674}
Y.~Avishai and Y.~B. Band, ``One-dimensional density of states and the phase of
  the transmission amplitude,''
  \href{http://dx.doi.org/10.1103/PhysRevB.32.2674}{{\em Phys. Rev. B}
  {\bfseries 32} (Aug, 1985) 2674--2676}.
  \url{https://link.aps.org/doi/10.1103/PhysRevB.32.2674}.

\bibitem{Hawking:1975vcx}
S.~W. Hawking, ``{Particle Creation by Black Holes},''
  \href{http://dx.doi.org/10.1007/BF02345020}{{\em Commun. Math. Phys.}
  {\bfseries 43} (1975) 199--220}. [Erratum: Commun.Math.Phys. 46, 206 (1976)].

\bibitem{PhysRev.187.345}
R.~Dashen, S.-k. Ma, and H.~J. Bernstein, ``S-matrix formulation of statistical
  mechanics,'' \href{http://dx.doi.org/10.1103/PhysRev.187.345}{{\em Phys.
  Rev.} {\bfseries 187} (Nov, 1969) 345--370}.
  \url{https://link.aps.org/doi/10.1103/PhysRev.187.345}.

\bibitem{Witten:2021jzq}
E.~Witten, ``{Why Does Quantum Field Theory In Curved Spacetime Make Sense? And
  What Happens To The Algebra of Observables In The Thermodynamic Limit?},''
  \href{http://arxiv.org/abs/2112.11614}{{\ttfamily arXiv:2112.11614
  [hep-th]}}.

\bibitem{Berti:2009kk}
E.~Berti, V.~Cardoso, and A.~O. Starinets, ``{Quasinormal modes of black holes
  and black branes},''
  \href{http://dx.doi.org/10.1088/0264-9381/26/16/163001}{{\em Class. Quant.
  Grav.} {\bfseries 26} (2009) 163001},
  \href{http://arxiv.org/abs/0905.2975}{{\ttfamily arXiv:0905.2975 [gr-qc]}}.

\bibitem{Hadar:2022xag}
S.~Hadar, D.~Kapec, A.~Lupsasca, and A.~Strominger, ``{Holography of the Photon
  Ring},'' \href{http://arxiv.org/abs/2205.05064}{{\ttfamily arXiv:2205.05064
  [gr-qc]}}.

\bibitem{Konoplya:2011qq}
R.~A. Konoplya and A.~Zhidenko, ``{Quasinormal modes of black holes: From
  astrophysics to string theory},''
  \href{http://dx.doi.org/10.1103/RevModPhys.83.793}{{\em Rev. Mod. Phys.}
  {\bfseries 83} (2011) 793--836},
  \href{http://arxiv.org/abs/1102.4014}{{\ttfamily arXiv:1102.4014 [gr-qc]}}.

\bibitem{Horowitz:1999jd}
G.~T. Horowitz and V.~E. Hubeny, ``{Quasinormal modes of AdS black holes and
  the approach to thermal equilibrium},''
  \href{http://dx.doi.org/10.1103/PhysRevD.62.024027}{{\em Phys. Rev. D}
  {\bfseries 62} (2000) 024027},
  \href{http://arxiv.org/abs/hep-th/9909056}{{\ttfamily arXiv:hep-th/9909056}}.

\bibitem{Birmingham:2001pj}
D.~Birmingham, I.~Sachs, and S.~N. Solodukhin, ``{Conformal field theory
  interpretation of black hole quasinormal modes},''
  \href{http://dx.doi.org/10.1103/PhysRevLett.88.151301}{{\em Phys. Rev. Lett.}
  {\bfseries 88} (2002) 151301},
  \href{http://arxiv.org/abs/hep-th/0112055}{{\ttfamily arXiv:hep-th/0112055}}.

\bibitem{Sun:2020ame}
Z.~Sun, ``{AdS one-loop partition functions from bulk and edge characters},''
  \href{http://dx.doi.org/10.1007/JHEP12(2021)064}{{\em JHEP} {\bfseries 12}
  (2021) 064}, \href{http://arxiv.org/abs/2010.15826}{{\ttfamily
  arXiv:2010.15826 [hep-th]}}.

\bibitem{Giombi:2013fka}
S.~Giombi and I.~R. Klebanov, ``{One Loop Tests of Higher Spin AdS/CFT},''
  \href{http://dx.doi.org/10.1007/JHEP12(2013)068}{{\em JHEP} {\bfseries 12}
  (2013) 068}, \href{http://arxiv.org/abs/1308.2337}{{\ttfamily arXiv:1308.2337
  [hep-th]}}.

\bibitem{Giombi:2014iua}
S.~Giombi, I.~R. Klebanov, and B.~R. Safdi, ``{Higher Spin AdS$_{d+1}$/CFT$_d$
  at One Loop},'' \href{http://dx.doi.org/10.1103/PhysRevD.89.084004}{{\em
  Phys. Rev. D} {\bfseries 89} no.~8, (2014) 084004},
  \href{http://arxiv.org/abs/1401.0825}{{\ttfamily arXiv:1401.0825 [hep-th]}}.

\bibitem{Giombi:2016pvg}
S.~Giombi, I.~R. Klebanov, and Z.~M. Tan, ``{The ABC of Higher-Spin AdS/CFT},''
  \href{http://dx.doi.org/10.3390/universe4010018}{{\em Universe} {\bfseries 4}
  no.~1, (2018) 18}, \href{http://arxiv.org/abs/1608.07611}{{\ttfamily
  arXiv:1608.07611 [hep-th]}}.

\bibitem{Gunaydin:2016amv}
M.~G\"unaydin, E.~D. Skvortsov, and T.~Tran, ``{Exceptional $F(4)$ higher-spin
  theory in AdS$_{6}$ at one-loop and other tests of duality},''
  \href{http://dx.doi.org/10.1007/JHEP11(2016)168}{{\em JHEP} {\bfseries 11}
  (2016) 168}, \href{http://arxiv.org/abs/1608.07582}{{\ttfamily
  arXiv:1608.07582 [hep-th]}}.

\bibitem{Grewal:2021bsu}
M.~Grewal and K.~Parmentier, ``{Characters, quasinormal modes, and Schwinger
  pairs in dS$_{2}$ with flux},''
  \href{http://dx.doi.org/10.1007/JHEP03(2022)165}{{\em JHEP} {\bfseries 03}
  (2022) 165}, \href{http://arxiv.org/abs/2112.07630}{{\ttfamily
  arXiv:2112.07630 [hep-th]}}.

\bibitem{Dolan:2005wy}
F.~A. Dolan, ``{Character formulae and partition functions in higher
  dimensional conformal field theory},''
  \href{http://dx.doi.org/10.1063/1.2196241}{{\em J. Math. Phys.} {\bfseries
  47} (2006) 062303}, \href{http://arxiv.org/abs/hep-th/0508031}{{\ttfamily
  arXiv:hep-th/0508031}}.

\bibitem{Giombi:2008vd}
S.~Giombi, A.~Maloney, and X.~Yin, ``{One-loop Partition Functions of 3D
  Gravity},'' \href{http://dx.doi.org/10.1088/1126-6708/2008/08/007}{{\em JHEP}
  {\bfseries 08} (2008) 007}, \href{http://arxiv.org/abs/0804.1773}{{\ttfamily
  arXiv:0804.1773 [hep-th]}}.

\end{thebibliography}\endgroup

\end{document}